\documentclass[a4paper]{aa}
\usepackage{graphicx}
\usepackage{txfonts}

\usepackage[english]{babel}

\usepackage{natbib}
\bibpunct{(}{)}{;}{a}{}{,}

\usepackage{color}

\renewcommand{\imath}{\mathrm{i}}

\begin{document} 

   \title{Depolarization of synchrotron radiation in a multilayer magneto-ionic medium}
\titlerunning{Depolarization of synchrotron radiation}
   \author{C. Shneider \inst{1} \and M. Haverkorn \inst{$2,1$} \and A. Fletcher \inst{3} \and A. Shukurov \inst{3} }

    \institute{Leiden Observatory, Leiden University,P.O. Box 9513, 2300 RA Leiden, The Netherlands \\ \email{shneider@strw.leidenuniv.nl} \and  Department of Astrophysics/IMAPP, Radboud University Nijmegen, P.O. Box 9010, 6500 GL Nijmegen, The Netherlands \and School of Mathematics and Statistics, Newcastle University, Newcastle upon Tyne NE1 7RU, U.K.}

   \date{Accepted 04 April 2014} 
 
  \abstract{Depolarization of diffuse radio synchrotron emission is classified in terms of wavelength-independent and wavelength-dependent depolarization in the context of regular magnetic fields and of both isotropic and anisotropic turbulent magnetic fields. Previous analytical formulas for depolarization due to differential Faraday rotation are extended to include internal Faraday dispersion concomitantly, for a multilayer synchrotron emitting and Faraday rotating magneto-ionic medium. In particular, depolarization equations for a two- and three-layer system (disk-halo, halo-disk-halo) are explicitly derived. 
To both serve as a `user's guide' to the theoretical machinery and as an approach for disentangling line-of-sight depolarization contributions in face-on galaxies, the analytical framework is applied to data from a small region in the face-on grand-design spiral galaxy M51. 
The effectiveness of the multiwavelength observations in constraining the pool of physical depolarization scenarios is illustrated for a two- and three-layer model along with a Faraday screen system for an observationally motivated magnetic field configuration.}  


   \maketitle

\section{Introduction}

 Depolarization of linearly polarized synchrotron radiation combined with multiwavelength observations is a powerful diagnostic for probing the constituents of the diffuse interstellar medium (ISM) in galaxies. The medium may be either synchrotron-emitting and Faraday-rotating or only Faraday-rotating (a Faraday screen) depending on whether cosmic ray electrons occur conjointly with thermal electrons and magnetic fields. Magnetic fields encompass regular (mean) fields, which are ordered and coherent on large scales and turbulent fields on small scales. 
Turbulent fields are further classified as isotropic or anisotropic. An alternative definition of anisotropy in terms of field striation may be found in \citet{farrar}. 
The three distinct components of the magnetic field - regular, turbulent isotropic, and turbulent anisotropic - contribute differently to the three observables of total synchrotron intensity (I), polarized synchrotron intensity (PI), and the Faraday rotation measure (RM) as discussed in \citet{tesse,farrar} (see Fig.1 of \citet{tesse} for an illustration). 

The study of depolarization signatures in synchrotron radiation has its origins in the suggestion by \citet{Alfven_Herlof} that cosmic radio waves result from relativistic electrons spiralling in magnetic fields. 
For an overview of observational tracers of galactic magnetic fields, see \citet{Zweibel_Heiles}.

In the context of nearby spiral galaxies, the basic results concerning polarization and Faraday effects stem from the seminal work of \citet{bJburn} who considered wavelength-dependent depolarization contributions from regular and \emph{isotropic} turbulent magnetic fields to describe the distribution of polarized radiation along the line of sight. 
Depolarization of synchrotron radiation by anisotropic magnetic fields and the effects of the magneto-ionic medium on the propogation of radio waves had already been described by \citet{ginzburg_syrovatskii}. 
In particular, \citet{korchakov} had arrived at wavelength-independent analytical formulas connecting the degree of polarization to the degree of regularity of the field for the presence of an \emph{anisotropic} magnetic field superimposed on a regular magnetic field as in the spiral arms of the Galaxy.
In their introduction, \citet{sokol,sokol_erratum} provide a concise summary of works on applications of depolarization laws to characterize magnetic fields in radio galaxies, jets, and other radio sources. 
\citet{bJburn} considered the case of a symmetric, single-layer uniform slab with constant emissivity and Faraday rotation per unit line of sight (for a review of several other models see \citet{Gardner66}).

In the sole presence of regular magnetic fields permeating the (Burn) slab, the polarization angle is a linear function of the square of the wavelength, and the degree of polarization follows the (Burn) depolarization (sinc) function. The Galactic foreground was modeled as a Burn slab in the work of \citet{brentjens05}.
When an isotropic Gaussian random magnetic field is also present the Burn depolarization formula is modified to include internal Faraday dispersion (IFD), with dispersion scaling with the quartic power of the wavelength. As noted by \citet{sokol}, a factor of `2' was missed in the dispersion formula.
Moreover, Faraday dispersion in an external screen was also examined and received criticism from \citet{tribble91} who modified this result to scale with the quadratic power of the wavelength since the dispersion would cause the spatial correlation length of the polarized emission to decrease with increasing wavelength until it would drop below the size of the turbulent cells (see also \citet{sokol}). 
\citet{bJburn} also considered wavelength-independent depolarization arising from variations in polarization angle by the presence of isotropic random magnetic fields. 
This led to the expression for the degree of polarization in terms of the ratio of energy densities of the regular and random magnetic fields as 
\begin{equation} \label{Burn_angle_var}
 \frac{p_{obs}}{p_{max}} \, = \, \frac{B^2_{u}}{B^2_{u} + B^2_{r}}, \nonumber \\
\end{equation}
which was corrected by \citet{Heiles96} to
\begin{equation} \label{Heiles_angle_var}
 \frac{p_{obs}}{p_{max}} \, = \, \frac{B^2_{u}}{B^2_{u} + \frac{2}{3}B^2_{r}}, \nonumber \\
\end{equation}
for a face-on spiral galaxy. 
Here, $p_{obs}$ and $p_{max}$ are the observed and maximum degrees of polarization, and $B_u$ and $B_r$ denote the uniform (regular) and random (isotropic turbulent) magnetic fields, respectively.

The work of \citet{sokol} generalizes the results of \citet{bJburn} to describe more complex lines of sight in which magnetic field reversals occur and which pass though a multilayer magneto-ionic medium as characteristic of spiral galaxies. Emissivity and Faraday rotation are no longer constant and may arise from cosmic ray electrons and thermal electrons with differing extents along the line of sight. 
These authors consider the cases of a symmetric nonuniform slab, an asymmetric slab, and a multilayer slab and show that the polarization angle is no longer a linear function of the wavelength squared in all of these contexts. 
Additionally, formulas for wavelength-independent depolarization arising from isotropic turbulent and anisotropic turbulent magnetic fields are derived using the rms value for the turbulent magnetic field strength.

We base our method on the multilayer slab approach but now include the simultaneous action of differential Faraday rotation (DFR) and IFD in each layer of a two- or three-layer magneto-ionic medium. An explicit analytical formula for polarization arising from a three-layer medium is provided. We also combine wavelength-dependent and wavelength-independent effects in this framework and allow for regular, isotropic random, and anisotropic random magnetic fields. 
To the authors' knowledge, this is also the first specific application (in modeling) of the analytical work done on anisotropic fields.
	
This multilayer approach is intended for modeling nearly face-on galaxies where it is difficult to disentangle the signal from the disk and halo.   
We apply the developed theoretical machinery to the face-on, grand-design spiral galaxy M51, which lends itself to a decomposition into a disk and a halo thanks to its small angle of inclination. 

In this paper, we lay the foundations for an improved physical modeling of the galaxy, building on previous works \citep{berk,fletcher11} by taking depolarizing effects into account directly, thus enabling a statistical comparison with polarization maps at each observing wavelength. 
In a follow-up paper, we will apply the method to constrain both regular and turbulent field strengths in M51 (Shneider et al. in prep., Paper II).

\section{Method} \label{method}
  
\subsection{Regular, isotropic turblent, and anisotropic turbulent} \label{regisoanisorand}

  We model a nearly face-on spiral galaxy with a disk and a halo. The multilayer decomposition along the line of sight is performed explicitly for a two- (disk-halo) and three- (halo-disk-halo, with the far and near sides of the halo being identical) layer system, in order to examine the depolarization contribution of the side of the halo farthest from the observer. Constant strength regular and turbulent magnetic fields along with a constant cosmic ray density $n_{\text{cr}}$ as well as a constant thermal electron density $n_\text{e}$ serve as independent input for the disk and halo. 
The effects of wavelength-independent and wavelength-dependent depolarization are directly traced by the normalized degree of polarization that describes the degree to which the measured polarized signal deviates from its intrinsic value. Several depolarization mechanisms are in play in the medium. We focus on the main ones for our modeling and discuss these separately.


 The total field is comprised of a regular and fluctuating (turbulent) part and is given by $\boldsymbol{B}  = \overline{\boldsymbol{B}} + \boldsymbol{b}$, where the over-bar notation has been adopted to denote the mean field. The fluctuating part is described by a three dimensional turbulent vector field $\boldsymbol{b}$ which is a random variable, with cylindrical components $b_r,\, b_{\phi}, \, b_z$ (in the galaxy plane) and whose standard deviation is similarly $\sigma_r,\,\sigma_\phi, \, \sigma_z$. A correlation between the transverse $b_\perp$ and longitudinal $b_z$ components of the turbulent magnetic field $\boldsymbol{b}$ arises from the solenoidality or divergence free condition $\nabla \cdot \boldsymbol{b} = 0$. It is assumed that the effect of such a correlation is negligible, thereby allowing for these components to be treated as uncorrelated \citep{sokol}. 

As soon as turbulent magnetic fields appear in the description, all related quantities have to be addressed through an expectation value given by a volume average over the random magnetic fluctuations in the source of synchrotron radiation. Since volume averaging will be equal to ensemble averaging in our treatment, the self consistency of the above representation for the total magnetic field may be obtained by \emph{ensemble} averaging both sides and noting that $\boldsymbol{b}$ and its components are random variables with zero mean. Hence, $\overline{\boldsymbol{B}}$ is also an ensemble average of the total field $\boldsymbol{B}$.
Upon including the three dimensional turbulent magnetic field $\boldsymbol{b}$ and assuming the standard scaling of emissivity with the square of the perpendicular component of the total magnetic field, $\varepsilon \propto B^2_\perp$, it is the expectation values of $\left\langle B_k \right\rangle = \overline{B}_k$ and $\left\langle B^2_k \right\rangle = \overline{B}^2_k \, + \, \sigma^2_k$ where $\sigma$ denotes the respective standard deviation with $k = \lbrace x,y,z \rbrace$ and $\langle \ldots \rangle$ represent expectation values or ensemble averages, which feature in equations describing depolarization. Please consult Appendix A for a more detailed explanation and an alternative scaling based on the equipartition assumption. 

 For isotropy, $\sigma_r = \sigma_{\phi} = \sigma_z = \sigma$. We include anisotropy caused by compression along spiral arms and by shear from differential rotation and assume it to have the form 
\begin{equation} \label{aniso_form_used}
\sigma^2_{\phi} = \alpha \, \sigma^2_r, \hspace{1mm} \sigma_r = \sigma_z, \\
\end{equation}
with $\alpha > 1$. Isotropy may be seen as the case where $\alpha = 1$. We emphasize that the above relations for isotropy and anisotropy, characterized by $\alpha$, are relations between the square of the standard deviation or variance of the components of $\boldsymbol{b}$ and \emph{not} among components of $\boldsymbol{b}$ itself.

\subsection{Projection from galaxy-plane to sky-plane coordinates} \label{galtoskyplane} 

 The total magnetic field and the intrinsic polarization angle of synchrotron radiation must be projected from the galaxy-plane onto the sky-plane. 
 For the regular disk and halo fields, the transformation from galaxy-plane cylindrical polar coordinates to sky-plane Cartesian coordinates proceeds with the introduction of two Cartesian reference frames, one with its origin at M51's center and the second in the sky-plane, with the $x$-axis of both frames pointing to the northern end of the major axis, and is given as \citep{berk}
 \begin{align}
  \overline B_x & = B_r \cos(\phi) \, - \, B_{\phi} \sin(\phi),				\nonumber \\  
  \overline B_y & = \left[ B_r \sin(\phi) \, + \, B_{\phi} \cos(\phi) \right] \cos(l) + B_z \sin(l), \nonumber \\  
  \overline B_{||} &=  - \left[ B_r \sin(\phi) \, + \, B_{\phi} \cos(\phi) \right] \sin(l) + B_z \cos(l),  \nonumber 	  
 \end{align}
where $l$ is the inclination angle and $||$ denotes a component of the field parallel to the line of sight.

 The random fields, represented by their standard deviations, transform to the sky-plane as
 \begin{align} \label{berkcoordtransf_turb}
 \sigma^2_x  &= \left \langle \left[ b_r \cos(\phi) - b_{\phi} \sin(\phi) \right]^2 \right \rangle \nonumber \\
			 &= \sigma^2_r \cos^2(\phi) + \sigma^2_{\phi} \sin^2(\phi),  \nonumber \\ 
 \sigma^2_y  &= \left \langle \left\{ \left[ b_r \sin(\phi) + b_{\phi} \cos(\phi) \right] \cos(l) + b_z \sin(l) \right\}^2 \right \rangle \nonumber \\
			 &= \left[ \sigma^2_r \sin^2(\phi) + \sigma^2_{\phi} \cos^2(\phi) \right] \cos^2(l) + \sigma^2_z \sin^2(l), \nonumber \\
 \sigma^2_{||}  &= \left \langle \left\{ -\left[ b_r \sin(\phi) + b_{\phi} \cos(\phi) \right] \sin(l) + b_z \cos(l) \right\}^2 \right \rangle \nonumber \\
	      		 &= \left[ \sigma^2_r \sin^2(\phi) + \sigma^2_{\phi} \cos^2(\phi) \right] \sin^2(l) + \sigma^2_z \cos^2(l). 	 		
 \end{align}
 It follows from Eqs.~\eqref{aniso_form_used} and~\eqref{berkcoordtransf_turb} that anisotropy is given by 
 \begin{align} \label{sigma_eq}
 \sigma^2_x &= \sigma^2_r \left[ \cos^2(\phi) + \alpha \sin^2(\phi) \right], \nonumber \\ 
 \sigma^2_y &= \sigma^2_r \left\{ \left[ \sin^2(\phi) + \alpha \cos^2(\phi) \right] \cos^2(l) + \sin^2(l) \right\}, \nonumber \\ 
 \sigma^2_{||} &= \sigma^2_r \left\{ \left[ \sin^2(\phi) + \alpha \cos^2(\phi) \right] \sin^2(l) + \cos^2(l) \right\}.  
 \end{align}

 The intrinsic polarization angle in the presence of regular fields only is given by \citep{sokol} 
  \begin{equation*} \label{psisubi}
  \psi_{0} = \tfrac{1}{2}\pi \, + \, \arctan \left( \overline B_{y} / \overline B_{x} \right)		
  \end{equation*}
which acquires an additional term under projection to the sky-plane to \citep{berk} 
  \begin{equation} \label{psisubi_coordtransform}
  \psi_{0} = \tfrac{1}{2}\pi \, - \, \arctan \left[\cos(l) \tan(\phi) \right] \, + \, \arctan \left( \overline B_{y} / \overline B_{x} \right).	
  \end{equation}

 With the inclusion of turbulent magnetic fields, the last term in the above equation is modified and the intrinsic angle becomes (see \citet{sokol} and Appendix A of this paper for a derivation of this modification) 
 \begin{equation} \label{psi_gen}
 \left \langle \psi_{0} \right \rangle = \tfrac{1}{2}\pi \, - \, \arctan \left[\cos(l) \tan(\phi) \right] \, + \, \tfrac{1}{2} \arctan \left( \frac{2 \overline B_x \overline B_y}{\overline B^2_x - \overline B^2_y + \sigma^2_x - \sigma^2_y} \right) 
 \end{equation}
which reduces to Eq.~\eqref{psisubi_coordtransform} for the isotropic case. Hence, for both regular fields without any turbulence and for purely isotropic turbulence the same equation for the intrinsic angle applies.

\section{The complex polarization} \label{complexpolarization} 

 As a result of the assumption that the transverse and longitudinal components of the turbulent magnetic field are uncorrelated, both the emissivity and the intrinsic polarization angle become independent of the total Faraday depth which, consequently, leads to a decoupling of the wavelength-independent and wavelength-dependent effects, and the complex polarization $\mathcal{P}$ for the total magnetic field $\boldsymbol{B}$ may therefore be expressed, based on \citet{sokol}, as 
\begin{align} \label{complex_Pol_turb}
 \mathcal{P} &= \left( \int_V dV \, w(\boldsymbol{r}) \left \langle \varepsilon(\boldsymbol{r}) \right \rangle_{W \times h} \right)^{-1} \nonumber \\
& \times \, \int_V \, dV \, \mathcal{P}_0 \, \left \langle \varepsilon(\boldsymbol{r}) \right \rangle_{W \times h} \, \exp\left[{ \, 2 \imath \, \left( 0.81 \, \lambda^2 \int^{z_i}_z n_e \overline B_{||} \, dl'  \right) }\right] \nonumber \\
& \times \, \left \langle \exp\left[{ \, 2 \imath \, \left( 0.81 \, \lambda^2 \int^{z_i}_z n_e b_{||} \, dl'  \right) }\right] \right \rangle_{W \times h}
 \end{align}
where the intrinsic, complex polarization $\mathcal{P}_0$ is
 \begin{equation} \label{Pnaught}
 \mathcal{P}_0 = p_0 \, w(\boldsymbol{r}) \, \frac{\left \langle \varepsilon(\boldsymbol{r}) \, \exp\left[{ \, 2 \imath \, \psi_0(\boldsymbol{r})}\right] \right \rangle_{W \times h}}{\left \langle \varepsilon(\boldsymbol{r}) \right \rangle_{W \times h}}.
 \end{equation} 

The intrinsic degree of linear polarization of synchrotron radiation is taken as $p_0 = 0.70$. $w(\boldsymbol{r})$ is the beam profile function of coordinates in the sky-plane, $\varepsilon$ is the synchrotron emissivity, and the quantity inside the expectation value angular brackets in the numerator of Eq.~\eqref{Pnaught} is known as the complex emissivity. $\overline B_{||}$ and $b_{||}$ are the mean and random  magnetic field components along the line of sight ($\mu$G), $n_\text{e}$ is the volume density of thermal electrons ($\mbox{cm}^{-3}$), $\psi_0$ is the intrinsic value of the local polarization angle $\psi$ at position $\boldsymbol{r}$, and $\lambda$ is the observing wavelength (m). $\left \langle \ldots \right \rangle_{W \times h}$ denotes volume averaging in the synchrotron source, encompassed by the beam cylinder, where $W$ is the area covered by the telescope beam and $h$ is the extent encompassed by a slice within the beam cylinder which should be much smaller than the scale height of the constituents of the magneto-ionic medium. Coordinate $l'$ is measured in pc along the line of sight with positive direction pointing toward the observer with $z_i$ denoting the boundary of either a synchrotron emitting region or a Faraday screen closest to the observer. 
  
 The complex polarization is linked to the \emph{observable} polarization quantities, the Stokes parameters $I,Q,U$, as
 \begin{equation*} \label{complexpolarandstokes1}
\mathcal{P} = p \, \exp\left({2 \imath \, \Psi}\right) 
 \end{equation*} 
where
\begin{equation*} \label{complexpolarandstokes2}
p = \frac{PI}{I} = \frac{\sqrt{(Q^2+U^2)}}{I}  
\end{equation*}
and
\begin{equation*} \label{psieqn}
\Psi = \tfrac{1}{2}\arctan\left(\frac{U}{Q}\right). 
\end{equation*} 
$PI$ is the polarized synchrotron intensity with $p = \left|\mathcal{P}\right|$ the degree of polarization, and $Q$ and $U$ may be seen to be the real and imaginary parts of $\mathcal{P}$, respectively, normalized by the total synchrotron intensity $I = \int_V \, \varepsilon \, dV$ and $\Psi$ is the \emph{observed} polarization angle.

The following additional assumptions are used in the succeeding analysis of depolarization: 
\begin{enumerate}[i.]

\item The degree of polarization $p$ and the polarization angle $\psi$ are affected exclusively by depolarization mechanisms arising from the diffuse ISM \emph{within the galaxy itself}. \label{first}

\item A sufficiently large number of turbulent correlation cells for both $\varepsilon \exp\left(2 \imath \, \psi_0 \right)$ and $\varepsilon$, denoted as $N_W$, is encompassed by the telescope beam area in order to have \emph{deterministic} values for the complex polarization and, consequently, for the degree of polarization and polarization angle. \label{turb_corr_assum} 

\item The beam profile function is for a flat telescope beam profile with $w(\boldsymbol{r}) = 1$. \label{beamprofile}

\item The variation of parameters perpendicular to the line of sight is negligible within the telescope beam.  \label{varperpsight}

\item The expectation value of the intrinsic complex polarization $\left \langle \mathcal{P}_0 \right \rangle$ is not a function of the line of sight coordinate, where $\mathcal{P}_0$ is defined in Eq.~\eqref{Pnaught} above. In general, this assumption no longer holds if the equipartition assumption is invoked as the longitudinal component of the total field $B_{||}$ enters the scene and it may be a function of the line of sight coordinate (see Appendix~\ref{derivofpre_27_Sokol}). \label{zindepassumption}     

\end{enumerate}

For a multilayer system it may be shown by direct integration of Eq.~\eqref{complex_Pol_turb} along the line of sight $l$, with appropriate boundary conditions, that 
\begin{align}  
\hspace{0.5cm} \mathcal{P} &= \left( \sum\limits_{i=1}^N \, \left\langle\varepsilon_i\right\rangle \, L_i \right)^{-1} \times \, \sum\limits_{i=1}^N \, \left\langle\mathcal{P}_{0i}\right\rangle \, \left\langle\varepsilon_i\right\rangle \, \Bigg( \int^{L}_0  \exp\bigg\{ \nonumber \\ 
&{\int^{L}_z \, \left[ 2 \imath \, \big( 0.81 \, \lambda^2 \, n_{ei} \overline B_{||i} \big) \, - \, d_i \, \lambda^4 \, {\big( 0.81 \, \left\langle n_{ei} \right\rangle \, b_{||i} \big)}^2 \right] \, dl' }\bigg\} \, dl \, \Bigg)  \label{complex_Pol_turb_new} \\
	&= \sum\limits_{i=1}^N \, \left\langle\mathcal{P}_{0i}\right\rangle \, \frac{I_i}{I} \, \left[ \frac{1 - \exp{\, \left( -2 \, \sigma^2_{RM_i} \lambda^4 + 2 \, \imath \, R_i \lambda^2 \right) }}{2 \, \sigma^2_{RM_i} \lambda^4 - 2 \, \imath \, R_i \lambda^2} \right] \nonumber \\
& \times \, \exp\left[{2 \imath \, \left( \sum\limits_{j=i+1}^N \, R_j \, \lambda^2 \right)}\right], \label{complex_Pol_turb_new_final}
\end{align} 
where the per-layer total synchrotron emission $I_i$, the total Faraday depth\footnote{Faraday depth and Faraday rotation measure (RM) are equivalent when the observed polarization angle $\Psi$ is a linear function of $\lambda^2$ such as in a medium where synchrotron emission and Faraday rotation are separated. They differ only when this linearity no longer holds as for a medium with synchrotron emission and Faraday rotation mixed. A positive Faraday depth means that the magnetic field points toward the observer. See \citet{brentjens05} for further discussion.} $R_i$, and the dispersion of the intrinsic rotation measure (RM) within the volume of the telescope beam $\sigma_{RM_i}$ are respectively given as
  \begin{align} 
   I_i & = \varepsilon_i \, L_i, \nonumber \\
   R_i & = 0.81 \, n_{\text{e}i} \, \overline B_{||i} \, L_i,	\label{Rsubi} \\
   \sigma_{RM_i} &= 0.81 \, \left\langle n_{\text{e}i} \right\rangle \, b_{||i} \left(L_i \, d_i \right)^{1/2},   \label{sigma_RM}
   \end{align}
and where 
\begin{equation} \label{Pnaught_new}
\left\langle\mathcal{P}_{0i}\right\rangle = p_0 \frac{\left \langle \varepsilon_i \, \exp \left({2 \imath \, \psi_{0i}}\right) \right \rangle}{\left \langle \varepsilon_i \right \rangle} 
\end{equation}
is similarly given, as first introduced in Eq.~\eqref{Pnaught}, but now as a layer-dependent, averaged quantity.
The $\sigma_{RM}$ of Eq.~\eqref{sigma_RM} will be used in our modeling of wavelength-dependent depolarization due to isotropic and anisotropic turbulent magnetic fields in Section~\ref{intfdisp}. 
In so doing, we make the implicit assumption that $\sigma_{RM}$ may be taken as independent of observing angle as for a purely random magnetic field. 
From Eq.~\eqref{complex_Pol_turb_new_final} we observe that wavelength-independent depolarization contributions may be directly appended to the terms expressing wavelength-dependent depolarization as if they were effectively constants. 

The sum in Eqs.~\eqref{complex_Pol_turb_new} and~\eqref{complex_Pol_turb_new_final} is over independent, \emph{uniform} layers indexed by $i$ and $N$ is the total number of layers in the medium with the $N$th layer nearest the observer. $\psi_{0i}$ is the initial angle of polarization (rad), 
$L = \sum_{i} L_i$ is the total path length through the medium (pc), 
$I = \sum_{i} I_i$ is the total synchrotron intensity from all layers, and
$d_i$ is the diameter of a turbulent cell (pc) in a layer.
A constructive feature of the complex polarization $\mathcal{P}$ is that it is an \emph{additive} quantity; the total combined complex polarization from all layers is the sum of the complex polarizations arising in each layer weighted by the fractional synchrotron intensity $I_{i}/I$.

\section{Wavelength-independent depolarization} \label{waveinddep} 
 
 From Eq.~\eqref{Pnaught_new} we observe that wavelength-independent depolarization can only modify the intrinsic degree of polarization in the presence of turbulent magnetic fields.
It stems from a tangling of magnetic field lines in the emission region both along the line of sight and across the beam on all scales.
Denoting the isotropic, anisotropic, and isotropic with anisotropic instances of $\left( \left|\left\langle \mathcal{P}_{0i} \right\rangle\right|/p_0 \right)$) by $(W_I)_i$, $(W_A)_i$, and $(W_{AI})_i$, as well as a generic wavelength-independent depolarizing term by $W_i$, we have \citep{sokol}
 \begin{equation} \label{gen_form}  
 (W_A)_i \, = \, \left\{ \frac{\left \lbrack \left( \overline B^2_x - \overline B^2_y + \sigma^2_x - \sigma^2_y \right)^{2} + 4 \overline B^2_x  \overline B^2_y \right \rbrack^{1/2}}{\overline{B^2_\perp}} \right\}_i , 
 \end{equation}
where $\overline{B}^2_\perp = \overline B^2_x \,+ \, \overline B^2_y$ and $\overline{B^2_\perp} = \overline B^2_\perp \, + \, \sigma^2_x \, + \, \sigma^2_y $ (see Appendix~\ref{derivofpre_27_Sokol} for a derivation). The subscripted $i$ appears on the braces to indicate that all magnetic fields occurring in the equation are representative of a particular layer. 
Equation~\eqref{gen_form} reduces in the isotropic case to  
 \begin{equation} \label{iso_only}
 (W_I)_i \, = \, \left(\frac{\overline B^2_\perp}{\overline B^2_\perp + 2 \sigma^2} \right)_i . 
 \end{equation}
When both isotropic and anisotropic fields are present in a layer then 
 \begin{equation} \label{iso_and_aniso}
 (W_{AI})_i \, = \, \left(\frac{\overline B^2_\perp}{\overline B^2_\perp + 2 \sigma^2} \right)_i \, \underbrace{\left\{ \frac{\left \lbrack \left( \overline B^2_x - \overline B^2_y + \sigma^2_x - \sigma^2_y \right)^{2} + 4 \overline B^2_x  \overline B^2_y \right \rbrack^{1/2}}{\overline{B^2_\perp}} \right\}_i}_{\sigma_x \, \neq \, \sigma_y} .
 \end{equation}
With the occurrence of both isotropic and anisotropic turbulent magnetic fields in the same layer, there is consecutive depolarization by these fields as contained in Eq.~\eqref{iso_and_aniso}. The two turbulent fields are viewed as describing two spatially separate, bulk regions in the galaxy that do not interact. 

 In the context of a purely random field $\boldsymbol{B} = \boldsymbol{b}$, from Eq.~\eqref{gen_form} it is observed that complete depolarization may be avoided only with an \emph{anisotropic} random magnetic field
 \begin{equation} \label{aniso_turb_only} 
 (W_A)_i \, = \, \left( \frac{\left|\, \sigma^2_x - \sigma^2_y \,\right|}{\sigma^2_x + \sigma^2_y} \right)_i, \, \sigma_x \, \neq \, \sigma_y .
 \end{equation}
Equation~\eqref{aniso_turb_only} implies that the smaller the difference between $\sigma_x$ and $\sigma_y$, the nearer the turbulent field to being purely random, and the closer the signal to being completely depolarized. On the other hand, the greater the difference between the standard deviations, the weaker the contribution of wavelength-independent depolarization, and the closer the signal to its intrinsic degree of polarization. 
 In the absence of any random fields, $\sigma_k = 0$, and it is readily observed that there is no wavelength-independent depolarization contribution, with $\left| \left\langle\mathcal{P}_{0i}\right\rangle \right| = p_0$, in Eqs.~\eqref{gen_form} -~\eqref{iso_and_aniso}.

\section{Wavelength-dependent depolarization} \label{wavedepdepol}

\subsection{Differential Faraday rotation} \label{DFRsect}

 Differential Faraday rotation occurs when emission from different depths in the emitting layer, along the \emph{same} line of sight, experience different amounts of Faraday rotation due to the presence of \emph{regular} fields. For a regular field only, $\boldsymbol{B} = \overline{\boldsymbol{B}}$, Eq.~\eqref{complex_Pol_turb_new_final} becomes \citep{sokol}
\begin{equation} \label{complex_Pol_reg_new_final} 
{\mathcal{P}}_{\left(\boldsymbol{B} = \overline{\boldsymbol{B}}\right)} = p_0 \, \sum\limits_{i=1}^N \, \frac{I_i}{I} \, \frac{\sin \left(R_i \lambda^2\right)}{\left(R_i \lambda^2\right)} \, \exp\left[2 \imath \, \left( \psi_{0i} \, + \, \frac{R_i}{2} \lambda^2 \, + \sum\limits_{j=i+1}^N \, R_j \, \lambda^2 \right)\right].
\end{equation}
Equation~\eqref{complex_Pol_reg_new_final} shows that the polarized emission coming from a given layer has an initial degree of polarization determined by the Faraday depth in that layer and that the signal's intrinsic polarization angle undergoes Faraday rotation with $RM = R_i/2$ in the originating layer and $RM = R_j$ in each successive layer, which function as Faraday screens for the emission from layers deeper than themselves.
	
For the goal of this paper, the above equation is explicitly expanded to a two- and three-layer medium.
For a two-layer system, with a halo between the disk and observer, this is given by 			
 \begin{align} \label{2layer_dfr}
      \left(\frac{p}{p_{0}}\right)_{2layer} & = \Bigg| \frac{I_{d}}{I} \frac {\sin \left(R_d \lambda^2 \right)}{\left(R_d \lambda^2 \right)} e^{ 2 \imath \, \left[ \psi_{0d} \, + \, \left( \frac{R_{d}}{2} \, + \, R_h \right) \, \lambda^2 \right] } \nonumber \\
                 & \hspace{0.5cm} + \frac{I_{h}}{I} \frac {\sin \left(R_h \lambda^2 \right)}{\left(R_h \lambda^2 \right)} e^{ 2 \imath \, \left( \psi_{0h} \, + \, \frac{R_{h}}{2} \lambda^2 \right) }  \Bigg| \nonumber 	\\
		      & = \left\{ A_d^2 + A_h^2 + 2 \, A_d \, A_h \, \cos \left[ 2 \, \Delta \psi_{dh} + \left( R_d + R_h \right) \lambda^2 \right] \right\}^{1/2},	
 \end{align}
where 
  \begin{equation} \label{Asubi}
   A_i = \left(I_i / I \right) \frac {\sin \left(R_i \lambda^2 \right)}{\left(R_i \lambda^2 \right)} = \left(I_i / I \right) \, {\sin} \text{c} \left(R_i \, \lambda^2\right).  
  \end{equation}
The subscripts $i = d,h$ refer to the disk and halo, and $\Delta \psi_{dh} = \left\langle\psi_{0d}\right\rangle - \left\langle\psi_{0h}\right\rangle$ is the difference in the intrinsic angle of polarization between the disk and halo. 
Equation~\eqref{2layer_dfr}, in particular, is a typo-corrected form of the equation as it appears in \citet{sokol}, and it was derived in the work of \citet{Rebecca}.
The corresponding equation for a three-layer (halo-disk-halo) system, where the far and near sides of the halos are identical, is given by
 \begin{align} \label{3layer_dfr}
      \left(\frac{p}{p_{0}}\right)_{3layer}  &= \Bigg|  \frac{I_{h}}{I} \frac {\sin \left(R_h \lambda^2 \right)}{\left(R_h \lambda^2 \right)} \left\{ e^{2 \imath \, \left[ \psi_{0h} \, + \, \left( \frac{3 R_{h}}{2} \, + \, R_d \right) \, \lambda^2 \right] } \, + \, e^{2 \imath \, \left( \psi_{0h} \, + \, \frac{R_h}{2} \lambda^2 \right) } \right\} \nonumber \\
&\hspace{0.5cm} + \frac{I_{d}}{I} \frac {\sin \left(R_d \lambda^2 \right)}{\left(R_d \lambda^2 \right)} \, e^{ 2 \imath \, \left[ \psi_{0d} \, + \, \left( \frac{R_{d}}{2} \, + \, R_h \right) \, \lambda^2 \right] }  \Bigg| \nonumber 	\\
	& = \Bigg( 2 \, A_h^2 \, \bigg\{ 1 \, + \, \cos \left[ 2 \left( R_d + R_h \right) \lambda^2 \right] \bigg\} + A_d^2  \nonumber	\\
&\hspace{0.5cm} +  2 \, A_d \, A_h \, \bigg\{ \cos \left[ -2 \, \Delta \psi_{dh} + \left( R_d + R_h \right) \lambda^2 \right] \nonumber	\\
&\hspace{0.5cm} + \cos \left[ 2 \, \Delta \psi_{dh} + \left(R_d + R_h \right) \lambda^2 \right] \bigg\} \Bigg)^{1/2}.
 \end{align}

\subsection{Internal Faraday dispersion} \label{intfdisp}

Internal Faraday dispersion results from polarized signal undergoing different amounts of Faraday rotation both along the line of sight and across the telescope beam \emph{within} a region of synchrotron emission when the telescope beam encompasses many turbulent cells.  

For a purely random field, $\boldsymbol{B} = \boldsymbol{b}$, Eq.~\eqref{complex_Pol_turb_new_final} becomes
\begin{equation} \label{complex_Pol_turb_only_181113} 
{\mathcal{P}}_{\left(\boldsymbol{B} = \boldsymbol{b}\right)} = \sum\limits_{i=1}^N \, \left\langle\mathcal{P}_{0i}\right\rangle \, \frac{I_i}{I} \, \frac{\sinh \left(\sigma^2_{RM_i} \lambda^4 \right)}{\left(\sigma^2_{RM_i} \lambda^4 \right)} \, \exp{\left(- \, \sigma^2_{RM_i} \lambda^4 \right)}.  
\end{equation}
In contrast to DFR, the intrinsic polarization angle remains completely unaffected by any contributions to the phase from Faraday dispersion because such contributions by random fields are zero on average. 

 Upon comparing Eqs.~\eqref{complex_Pol_reg_new_final} and~\eqref{complex_Pol_turb_only_181113}, it is apparent that the $A_i$ in Eq.~\eqref{Asubi} has been modified to \citep{bJburn,sokol}
   \begin{align} 			
  \tilde A_i &= \left(I_i / I \right) \left[\frac{1 - \exp{\left(-2 \, \sigma^2_{RM_i} \lambda^4 \right)}}{2 \, \sigma^2_{RM_i} \lambda^4}\right]  \nonumber \\
	      &= \left(I_i / I \right) \frac{\sinh \left(\sigma^2_{RM_i} \lambda^4 \right)}{\left(\sigma^2_{RM_i} \lambda^4 \right)} \, \exp{\left(- \, \sigma^2_{RM_i} \lambda^4 \right)},  \nonumber
  \end{align}
and that Eqs.~\eqref{2layer_dfr} and~\eqref{3layer_dfr} are modified to
\begin{align}
\left(\frac{p}{p_0}\right)_{2layer} &=  (WA)_d \, \tilde{A}_d + (WA)_h \, \tilde{A}_h,   \nonumber \\
\left(\frac{p}{p_0}\right)_{3layer} &=  2 \, (WA)_h \,\tilde{A}_h + (WA)_d \, \tilde{A}_d.  \nonumber
\end{align}
 
A fundamental physical change has been effected; the sinc function with its non-monotonic, $\pi$-periodic zero-crossings in Eq.~\eqref{complex_Pol_reg_new_final} has now been replaced by a monotonically decreasing function of Faraday depth in Eq.~\eqref{complex_Pol_turb_only_181113} as the product of a \emph{hyperbolic} sinc function with an exponential decay.

\subsection{External Faraday dispersion} \label{Fscreen} 

 When polarized emission is modeled as arising exclusively from the disk, by having the halo devoid of any cosmic ray electrons, a two- and three-layer model approach to depolarization becomes degenerate since there is no longer a sum over depolarization terms but rather a single term that describes the Faraday depolarization contribution from the disk, together with the influence of the near halo (nearest to the observer) on the polarized signal. In particular, the far halo, coming from a three-layer model, would be completely dormant in terms of polarized signal. With only regular fields present in the halo, the halo contributes with just a Faraday rotating phase term that does not affect the degree of polarization.

With the inclusion of turbulent fields in the halo, the halo functions as a Faraday screen, contributing an external Faraday dispersion (EFD) term. \textit{External} refers to the turbulent fields between the observer and the source. Having both regular and turbulent magnetic fields present in the disk and halo  
entails having DFR and IFD in the disk, together with EFD in the halo, and yields 	
\begin{align} 
 \left(\frac{p}{p_0}\right)_{EFD} &= \Bigg| \frac{\left\langle\mathcal{P}_{0d}\right\rangle}{p_0} \, \left[ \frac{1 - \exp{\, \left( -2 \, \sigma^2_{RM_d} \lambda^4 + 2 \, \imath \, R_d \lambda^2 \right) }}{2 \, \sigma^2_{RM_d} \lambda^4 - 2 \, \imath \, R_d \lambda^2} \right] \nonumber \\ 
&\times \, \exp{\left[2 \imath \, \left( \psi_{0d} \, + \, R_h \lambda^2 \right) \, -2 \sigma^2_{RM_h} \lambda^4\right]} \, \Bigg| \nonumber \\
      &= W_d \, \left[ \frac{1 - 2 \, e^{-2 \, \sigma^2_{RM_d} \lambda^4} \cos{\left(2 \, R_d \lambda^2\right)} + e^{-4 \, \sigma^2_{RM_d} \lambda^4}}{\left(-2 \, \sigma^2_{RM_d} \lambda^4\right)^2 + \Big(2 \, R_d\lambda^2\Big)^2} \right] \nonumber \\ 
&\times \, \exp{\left(-2 \sigma^2_{RM_h} \lambda^4\right)} \label{Faradayscreen_sigma_halo_depol_terms_only}.
 \end{align}
A fractional synchrotron intensity term $I_d/I$ does not appear since all of the synchrotron emission stems from the disk (i.e., $I_d = I$).

For regular magnetic fields in the disk alone, along with turbulent magnetic fields in the halo, the equation is the natural reduction of Eq.~\eqref{Faradayscreen_sigma_halo_depol_terms_only} in this limit and is given by \citep{bJburn,sokol}
 \begin{align} 
\left(\frac{p}{p_0}\right)_{EFD} &= \Bigg| \frac {\sin \left(R_d \lambda^2 \right)}{\left(R_d \lambda^2 \right)} \, \exp\left[{2 \imath \, \left( \psi_{0d} \, + \, \frac{R_d}{2} \lambda^2 \, + \, R_h \lambda^2 \right) \, -2 \sigma^2_{RM_h} \lambda^4 }\right] \Bigg| \nonumber \\ 
&= \frac {\sin \left(R_d \lambda^2 \right)}{\left(R_d \lambda^2 \right)} \, \exp\left({-2 \sigma^2_{RM_h} \lambda^4}\right). \label{Faradayscreen_depol_terms_only}
 \end{align}

\subsection{Depolarization from DFR with IFD} \label{DirectSummation} 

  We derive equations for depolarization arising from IFD occurring concomitantly with DFR from Eq.~\eqref{complex_Pol_turb_new_final}. 
 For a two-layer system (with a halo between the disk and observer as in Eq.~\eqref{2layer_dfr}), this is given by
 \begin{align} \label{directsum_2layer}
    \left(\frac{p}{p_{0}}\right)_{2layer} = \Bigg| &  \frac{\left\langle\mathcal{P}_{0d}\right\rangle}{p_0} \, \frac{I_{d}}{I} \left[ \frac{1 - e^{\, \left( -2 \, \sigma^2_{RM_d} \lambda^4 + 2 \, \imath \, R_d \lambda^2 \right) }}{2 \, \sigma^2_{RM_d} \lambda^4 - 2 \, \imath \, R_d \lambda^2} \right] e^{2 \imath \left( \psi_{0d} \, + \, R_h \lambda^2 \right)} \nonumber \\
	& +  \frac{\left\langle\mathcal{P}_{0h}\right\rangle}{p_0} \, \frac{I_{h}}{I} \left[ \frac{1 - e^{\, \left( -2 \, \sigma^2_{RM_h} \lambda^4 + 2 \, \imath \, R_h \lambda^2 \right) }}{2 \, \sigma^2_{RM_h} \lambda^4 - 2 \, \imath \, R_h \lambda^2} \right] e^{2 \imath \psi_{0h}} \Bigg| 	\nonumber \\
	=  \Bigg\{  & W^2_d \, \left( \frac{I_d}{I} \right)^2 \left( \frac{1 - 2e^{-\Omega_d}\cos{C_d}+e^{-2\Omega_d}}{\Omega^2_d + C^2_d}\right) \nonumber \\
& +  W^2_h \, \left( \frac{I_h}{I} \right)^2 \left( \frac{1 - 2e^{-\Omega_h}\cos{C_h}+e^{-2 \Omega_h}}{\Omega^2_h + C^2_h}\right) 	\nonumber \\
& + W_d W_h \, \frac{I_d I_h}{I^2} \frac{2}{F^2 + G^2} \Bigg[ \left \lbrace F,G \right \rbrace \left( 2 \Delta \psi_{dh} + C_h \right) \nonumber \\
& + e^{-(\Omega_d \, + \, \Omega_h)} \left \lbrace F,G \right \rbrace \left( 2 \Delta \psi_{dh} + C_d \right) \nonumber \\
& - e^{- \Omega_d} \left \lbrace F,G \right \rbrace \left( 2 \Delta \psi_{dh} + C_d + C_h \right) \nonumber \\
& - e^{- \Omega_h}  \left \lbrace F,G \right \rbrace \left( 2 \Delta \psi_{dh} \right) \Bigg]  	\Bigg\}^{1/2},
 \end{align}	
where $\Omega_d = 2 \sigma^2_{RM_d} \lambda^4 $, $\Omega_h = 2 \sigma^2_{RM_h} \lambda^4$, $C_d = 2 R_d \lambda^2$, $C_h = 2 R_h \lambda^2$, $F = \Omega_d \Omega_h \,+ \,C_d C_h$, $G = \Omega_h C_d \,- \, \Omega_d C_h$. 
The operation $ \left \lbrace F, G \right \rbrace \left(a \right)$ is defined as $ \left \lbrace F, G \right \rbrace \left(a \right) = F\cos \left(a \right) \,-\, G\sin \left(a \right)$. 

 The corresponding equation for a three-layer system (with far and near halos identical as in Eq.~\eqref{3layer_dfr}) is given by
 \begin{align} \label{directsum_3layer}
    & \left(\frac{p}{p_{0}}\right)_{3layer} = \Bigg| \frac{\left\langle\mathcal{P}_{0h}\right\rangle}{p_0} \, \frac{I_{h}}{I} \left[ \frac{1 - e^{\, \left( -2 \, \sigma^2_{RM_h} \lambda^4 + 2 \, \imath \, R_h \lambda^2 \right) }}{2 \, \sigma^2_{RM_h} \lambda^4 - 2 \, \imath \, R_h \lambda^2} \right] \bigg\{ e^{2 \imath \, \left[ \psi_{0h} \, + \, \left( R_d \, + \, R_h \right) \lambda^2 \right]}  \nonumber \\
	& + e^{2 \imath \, \psi_{0h}} \bigg\} \, + \, \frac{\left\langle\mathcal{P}_{0d}\right\rangle}{p_0} \, \frac{I_{d}}{I} \left[ \frac{1 - e^{\, \left( -2 \, \sigma^2_{RM_d} \lambda^4 + 2 \, \imath \, R_d \lambda^2 \right) }}{2 \, \sigma^2_{RM_d} \lambda^4 - 2 \, \imath \, R_d \lambda^2} \right] e^{2 \imath \, \left( \psi_{0d} \, + \, R_h \lambda^2 \right)} \Bigg| 	\nonumber \\
 	&= \Bigg(  2 \, W^2_h \, \left(\frac{I_h}{I} \right)^2 \left\{ \frac{\left(1 - 2e^{-\Omega_h}\cos{D}+e^{-2\Omega_h}\right) \Big[ 1 + \cos \left( C_d + C_h \right) \Big]}{\Omega^2_h + C^2_h}\right\} \nonumber \\
& +  W^2_d \, \left(\frac{I_d}{I} \right)^2 \left( \frac{1 - 2e^{-\Omega_d}\cos{C}+e^{-2\Omega_d}}{\Omega^2_d + C^2_d}\right)	\nonumber \\
& + W_d W_h \, \frac{I_d I_h}{I^2} \frac{2}{F^2 + G^2} \Bigg\{ \left \lbrace F,-G \right \rbrace \left( -2 \Delta \psi_{dh} + C_d \right) \nonumber \\
&+ \left \lbrace F,G \right \rbrace \left( 2 \Delta \psi_{dh} + C_h \right) \nonumber \\
& + e^{-(\Omega_d \,+ \, \Omega_h)} \Big[ \left \lbrace F,G \right \rbrace \left( 2 \Delta \psi_{dh} + C_d \right) + \left \lbrace F,-G \right \rbrace \left( -2 \Delta \psi_{dh} + C_h \right) 	 \Big] \nonumber \\
& - e^{-\Omega_d} \Big[ \left \lbrace F,G \right \rbrace \left( 2 \Delta \psi_{dh} + C_d + C_h \right) + \left \lbrace F,-G \right \rbrace \left( -2 \Delta \psi_{dh} \right) \Big] \nonumber \\
& - e^{-\Omega_h} \Big[ \left \lbrace F,-G \right \rbrace \left( -2 \Delta \psi_{dh} + C_d + C_h \right) + \left \lbrace F,G \right \rbrace \left( 2 \Delta \psi_{dh} \right) \Big] \Bigg\} \Bigg)^{1/2}.
 \end{align}
The symmetry properties of these equations will be reserved for discussion in Appendix~\ref{eqndiscussion}. The above equations explicitly show the competition between the turbulent and regular fields with the $\sigma_{RM}$ and $R$ strictly characterizing exponential decay and periodicity.

\begin{figure}
\centering	
\includegraphics[width=\hsize]{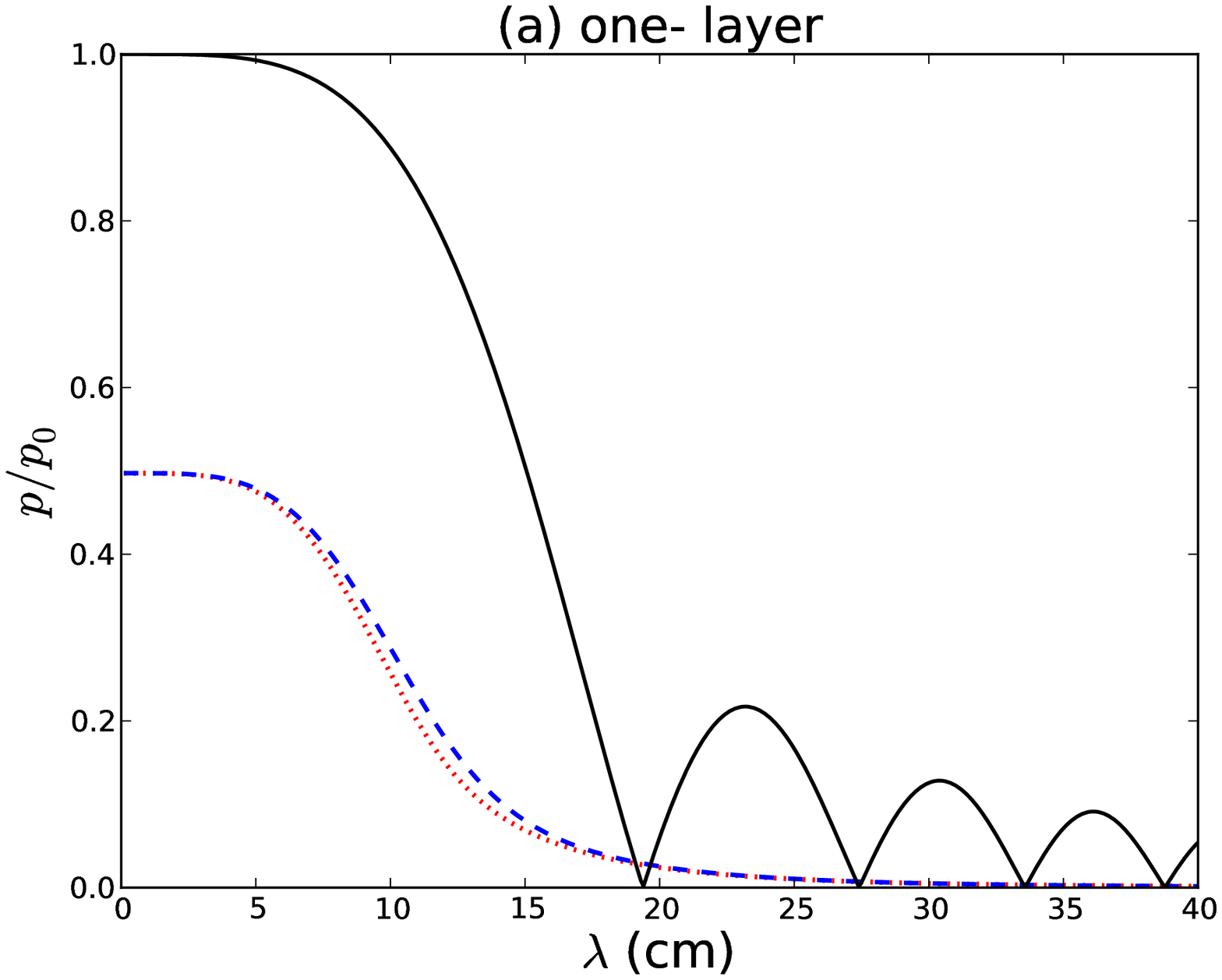}
\includegraphics[width=\hsize]{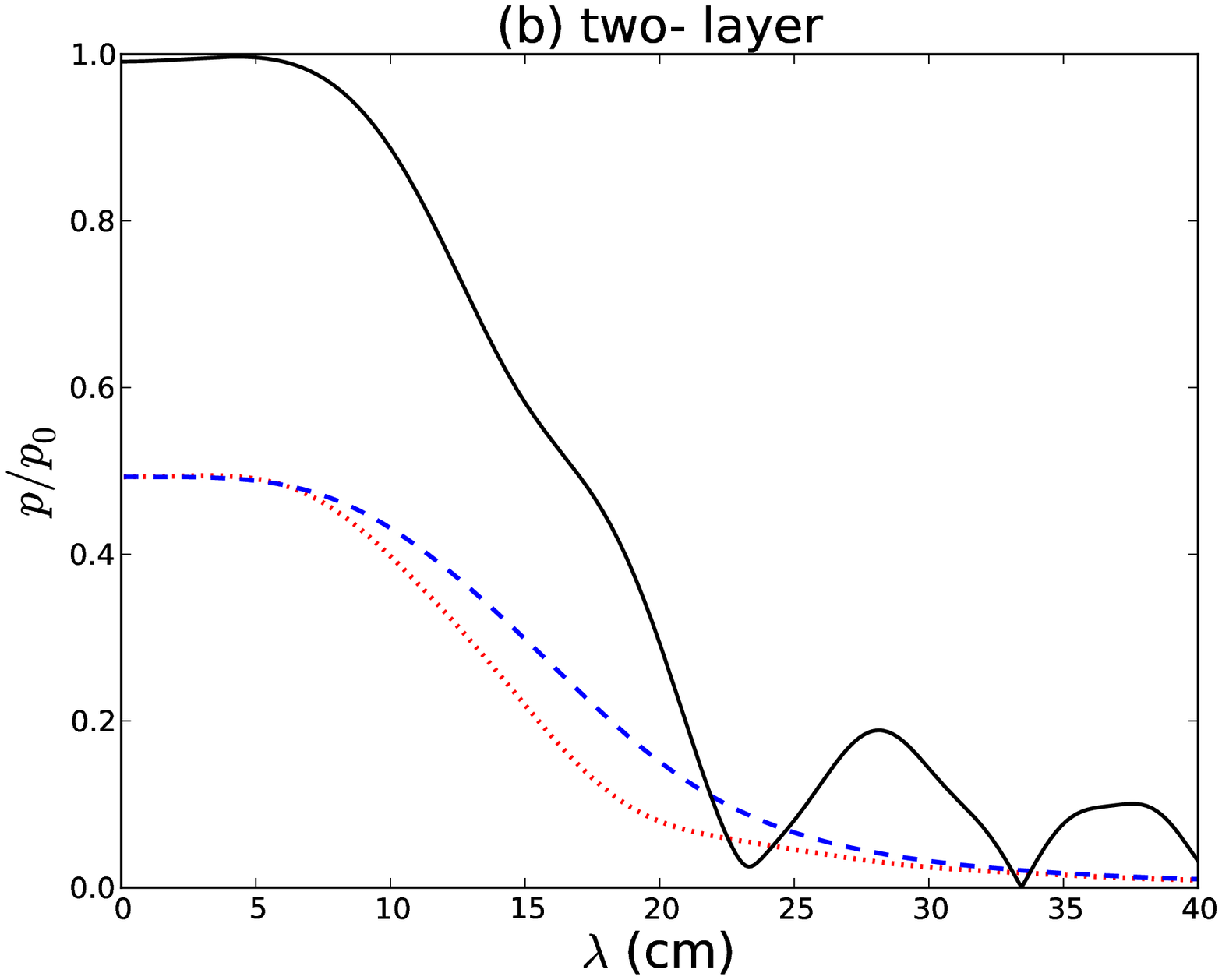} 
\includegraphics[width=\hsize]{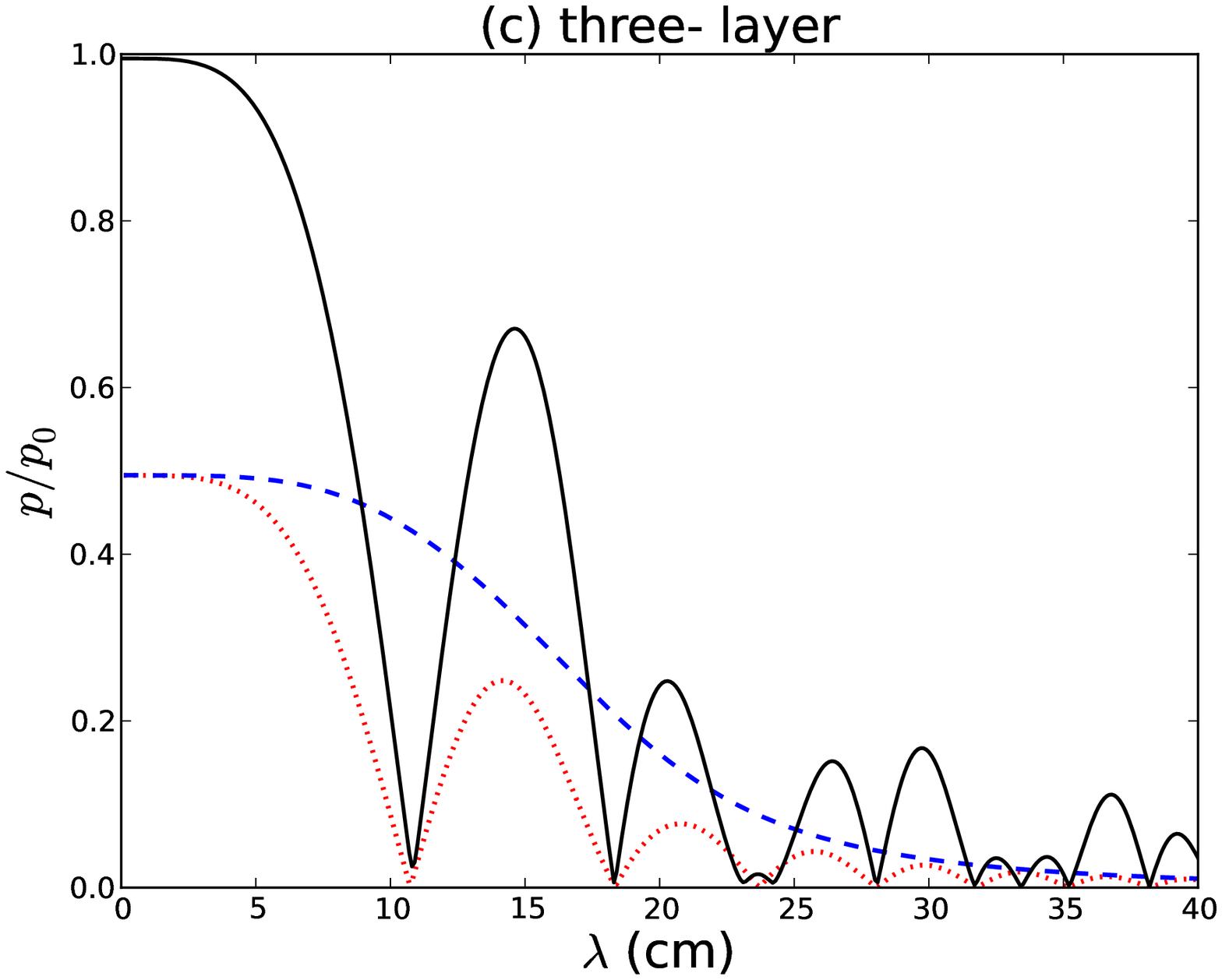} 
\caption{Normalized degree of polarization as a function of wavelength illustrated for a one-layer (a), two-layer (b), three-layer (c) system with characteristic profiles for DFR only (black solid), IFD only (blue dashed), and DFR with IFD (red dotted). 
A total isotropic turbulent magnetic field strength of $5 \, \mu$G together with a total regular magnetic field strength also of $5 \, \mu$G has been used in the disk and in the halo. 
The parameters of $n_{\text{e}},n_{\text{cr}},L,d,\alpha$ used in the construction of these plots are the same as those for the example bin of Section~\ref{applictoM51} and their values are reported in the bottom panel of Table~\ref{table:2}.}
\label{Fig4_factoriz123}
\end{figure}

Figure~\ref{Fig4_factoriz123} contains the depolarization profiles, with normalized degree of polarization plotted against wavelength, for a one-, two-, and three-layer magneto-ionic medium with DFR, IFD, and DFR with IFD. The wavelength-independent polarization has been assumed to be $0.5$ for illustration purposes. Its actual value should be fit to observations at a small enough wavelength to make wavelength-dependent depolarization effects negligible.
With an increasing number of magneto-ionic layers modeled, the DFR curve has complete depolarization occurring at progressively earlier wavelengths. 
Comparing the IFD curve for a single and multilayer medium reveals that the IFD curve persists at longer wavelengths and thus is less effective as a depolarizing mechanism in a multilayer medium. 
The `jagged' profile of the DFR curve in (b) relative to the smooth profile of (a) arises from there being two sinc functions with differing Faraday depths. For a three-layer system in (c), the halo sinc function alone determines the DFR curve thanks to the disk's small fractional synchrotron intensity, which accounts for the smoothness. 
Comparing the \citet{bJburn} and \citet{sokol} result for DFR with IFD in a one-layer uniform slab (a), represented by the sole presence of a disk, with that in a two-layer medium (b) given by a disk plus a halo reveals that the presence of a halo supports polarization at longer wavelengths. Similarly, DFR with IFD in a three-layer medium (c) with identical far and near sides of the halo undergoes a drastic change in profile, which more closely resembles a one-layer halo polarization profile.

\section{Modeling example: application to M51} \label{applictoM51} 
 
 We illustrate our method for the case of the nearby grand design spiral galaxy M51, with its high galactic latitude of $b = + \, 68.6^\circ$ and with an inclination angle $l = -20^{\circ}$. It is assumed that the observed emission is exclusively from M51 because of the high galactic latitude \citep{berk}. 
We use the \citet{fletcher11} model predictions of a two-dimensional regular magnetic field $\sum_m \boldsymbol{B}_m(r) \, \cos \left(m \, \phi - \beta_m\right)$ for both the disk and halo for a small region (a sector of radial size $1.2$ kpc and azimuthal extent $20^{\circ}$) of the galaxy. The turbulent magnetic field in the disk and halo is three dimensional. 
We compare the observed degrees of polarization at $\lambda\lambda\lambda \, 3.5, 6.2, 20.5$ cm with those expected from different models of the depolarization for this bin.

The regular disk and halo magnetic field configurations in cylindrical polar coordinates are
 \begin{align} \label{fletcherregflds}
 B_r & = B_0 \sin(p_0) + B_2 \sin(p_2)\cos(2 \phi - \beta_2), \nonumber \\
 B_{\phi} & =  B_0 \cos(p_0) + B_2 \cos(p_2)\cos(2 \phi - \beta_2), \nonumber \\
 B_z & = 0, \nonumber \\
 B_{\text{h}r} & = B_{\text{h}0}\sin(p_{\text{h}0}) + B_{\text{h}1}\sin(p_{\text{h}1})\cos(\phi - \beta_{\text{h}1}), \nonumber \\
 B_{\text{h}\phi} & = B_{\text{h}0}\cos(p_{\text{h}0}) + B_{\text{h}1}\cos(p_{\text{h}1})\cos(\phi - \beta_{\text{h}1}), \nonumber \\
 B_{\text{h}z} & = 0,
 \end{align}
where $p_m$ is the pitch angle of the total horizontal magnetic field, $\beta_m$ the azimuth at which the corresponding non $m=0$ mode is a maximum, and $h$ denotes the component of the halo field. The parameter values are given in Table~\ref{table:2}.
For anisotropic fields in the disk, $\alpha$ has been measured to be $1.83$ \citep{houde13} while for the halo anisotropic fields it is expected to be less than the disk value owing to weaker spiral density waves and differential rotation in the halo. In our model, the anisotropic factors for the disk and halo are $2.0$ and $1.5$, respectively.

\begin{table}
\caption{Parameters used to model the synchrotron polarization data for an example bin in M51 located in the innermost radial ring ($2.4 - 3.6$ kpc) at an azimuth centered on $100^{\circ}$.}	
\label{table:2}
\begin{center}
{\def\arraystretch{2}\tabcolsep=2pt
\begin{tabular}{l c c } 
\hline 
\hline 
 &  Disk & Halo \\ 
\hline
Mode ratios & $B_2/B_0 = (-33)/(-46)$ & $B_{\text{h}1}/B_{\text{h}0} = (76)/(23)$  \\ 
$p_m \, [^\circ$] & $p_0 = -20$, $p_2 = -12$ & $p_{\text{h}0} = -43$, $p_{\text{h}1} = -45$ \\
$\beta_m \, [^\circ$] & $\beta_2 = -8$ & $\beta_{\text{h}1} = 44$ \\
\hline
$n_{\text{e}}$ [cm$^{-3}$] & $0.11$ & $0.01$ \\
$n_{\text{cr}}$ [cm$^{-3}$]\tablefootmark{*} & const. & const. \\
$L$ [pc] & $800$ & $5000$ \\
$d$ [pc]\tablefootmark{**} & $40$ & $240$ \\
$\alpha$ & $2.0$ & $1.5$ \\ 
\hline
\end{tabular} 
}
\tablefoot{The fitted model parameters appearing in the upper panel for the regular magnetic field of Eq.~\eqref{fletcherregflds} are adopted from \citet{fletcher11} with central values reported only. 
The thermal electron density ($n_\text{e}$) and path length ($L$) for the disk and halo are gathered from \citet{fletcher11,berk}. \\
\tablefoottext{*}{The cosmic ray density is treated as a constant of proportionality between the synchrotron emissivity and the square of the total transverse magnetic field ($\mu \mbox{G}$) as $\varepsilon = c B^2_{\perp}$ with constant $c = 0.1$.} \\
\tablefoottext{**}{The turbulent cell size $d$ in the disk and halo is obtained from Eq.~\eqref{diam_turb_cell} with an RM dispersion $\sigma_{RM,D}$ fixed to the observed value of $15 \, \text{rad m}^{-2}$ within a telescope beam of linear diameter $D = 600$ pc. The rms value for the strength of the turbulent magnetic field along the line of sight $b^2_{||} = \sigma^2_{||}$ has been assumed, where the value for $\sigma^2_{||}$ is obtained via consideration~(\ref{rms}) with $\sigma^2_{\text{I}} = \sigma^2_{\text{A}} = 10 \, \mu \mbox{G}$ in the disk and $\sigma^2_{\text{I}} = \sigma^2_{\text{A}} = 3 \, \mu \mbox{G}$ in the halo. }
}
\end{center}
\end{table}

Table~\ref{table:1} shows all the possible model constituents. The model types are constructed based on the following considerations: 
\begin{enumerate}[i.]

\item The total synchrotron intensity (I) increases with the addition of turbulent fields since the ensemble average of the square of the transverse turbulent magnetic field component is non-zero $\left( \left \langle b_{\perp}^2  \right \rangle \neq 0 \right)$. This is also why the total intensity would be non-zero in the absence of any regular fields. \label{total_int} 
\item Root mean square (rms) values are used for the field strengths of the individual components of the turbulent magnetic fields in the disk and halo. 
The strength of an individual square component of the field $\sigma^2_k$ with $k = \lbrace x,y,|| \rbrace$ is obtained by substituting for $\sigma^2_r$ in Eq.~\eqref{sigma_eq} the normalized input isotropic $\sigma^2_{\text{I}}$ or anisotropic $\sigma^2_{\text{A}}$ field strength as $\sigma^2_r  = \sigma^2_{\text{I}} / 3$ for isotropy ($\alpha = 1$) and $\sigma^2_r  = \sigma^2_{\text{A}} / (2 \, + \, \alpha) $ for anisotropy. For completeness, $\sigma^2_{\phi} = \alpha \, \sigma^2_r$.
The anisotropic normalization factor in the galaxy plane is conserved upon projection to the sky plane. \label{rms} 

\item The diameter of a turbulent cell $d_i$ in the disk or halo is approximately given by \citep{fletcher11}
\begin{equation} \label{diam_turb_cell}
d_i \simeq \left[ \frac{D \,\sigma_{RM,D}}{ 0.81 \, \left \langle n_{\text{e}i} \right \rangle  \, b_{||i} \, (L_i)^{1/2}} \right]^{2/3}, 
\end{equation}
with $\sigma_{RM,D}$ denoting the RM dispersion observed within a telescope beam of a linear diameter $D = 600$ pc, and   
$\sigma_{RM,D}$ has been fixed to the observed value of $15 \, \text{rad m}^{-2}$. \label{rm_disp}

\end{enumerate}

\begin{figure}
\centering
\includegraphics[width=\hsize]{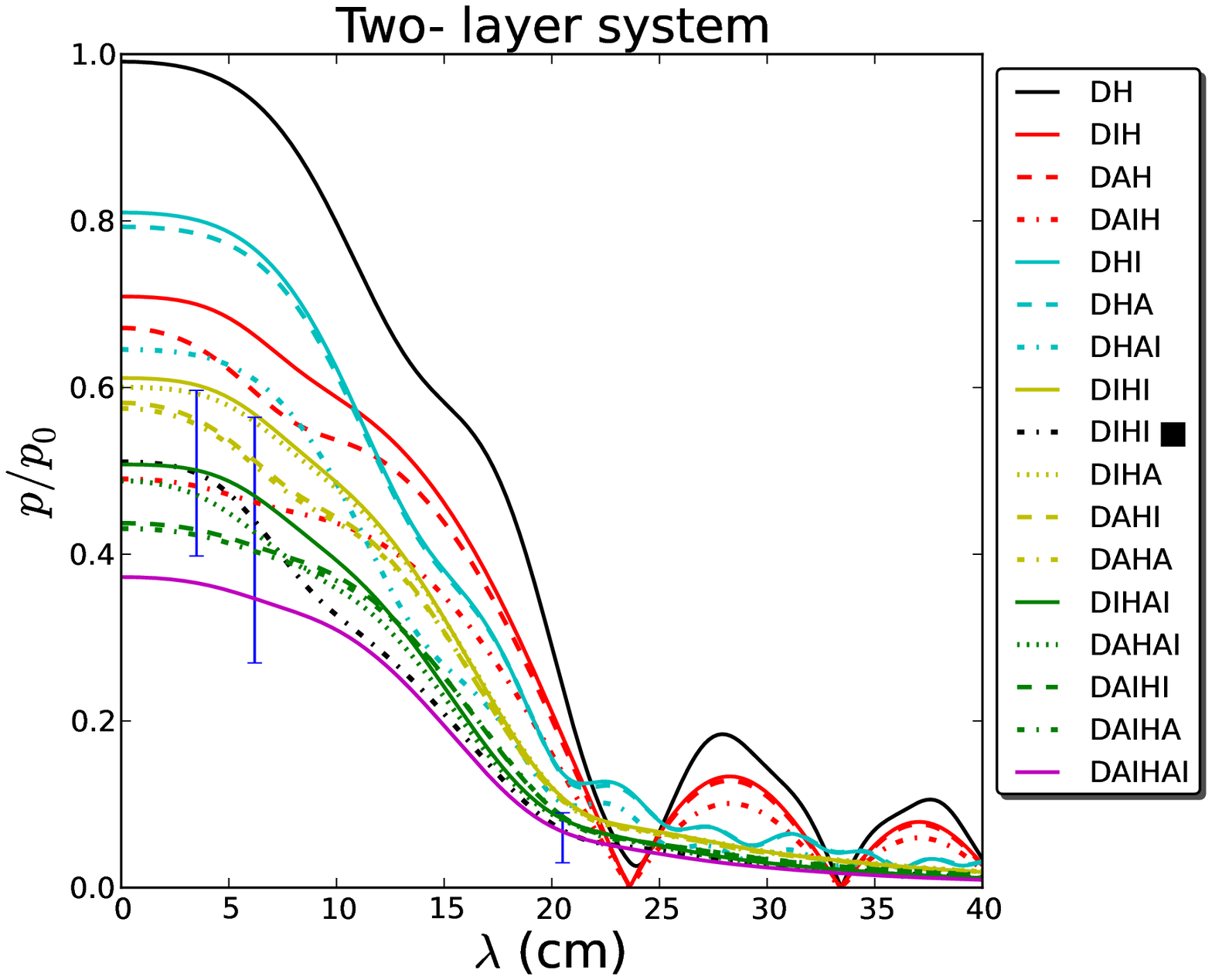} 
\caption{Normalized degree of polarization as a function of wavelength for a two-layer system description of M51. The measured polarization values for a sector with an azimuth centered at $100^{\circ}$ in radial ring 1 ($2.4 - 3.6$ kpc) at the three observing wavelengths $\lambda\lambda\lambda \, 3.5, 6.2, 20.5$ cm are displayed with error bars. 
All model profiles featured have been constructed from among the following set of magnetic fields: a total regular field strength of $5 \, \mu \mbox{G}$ in the disk and in the halo, an isotropic and anisotropic disk turbulent random field of $10 \, \mu \mbox{G}$ each, and an isotropic and anisotropic halo turbulent random field of $3 \, \mu \mbox{G}$ each.
Please consult Table~\ref{table:1} for nomenclature and description of the model types appearing in the legend.}
\label{Fig1_2layers}
\end{figure}

\begin{figure}
\centering
\includegraphics[width=\hsize]{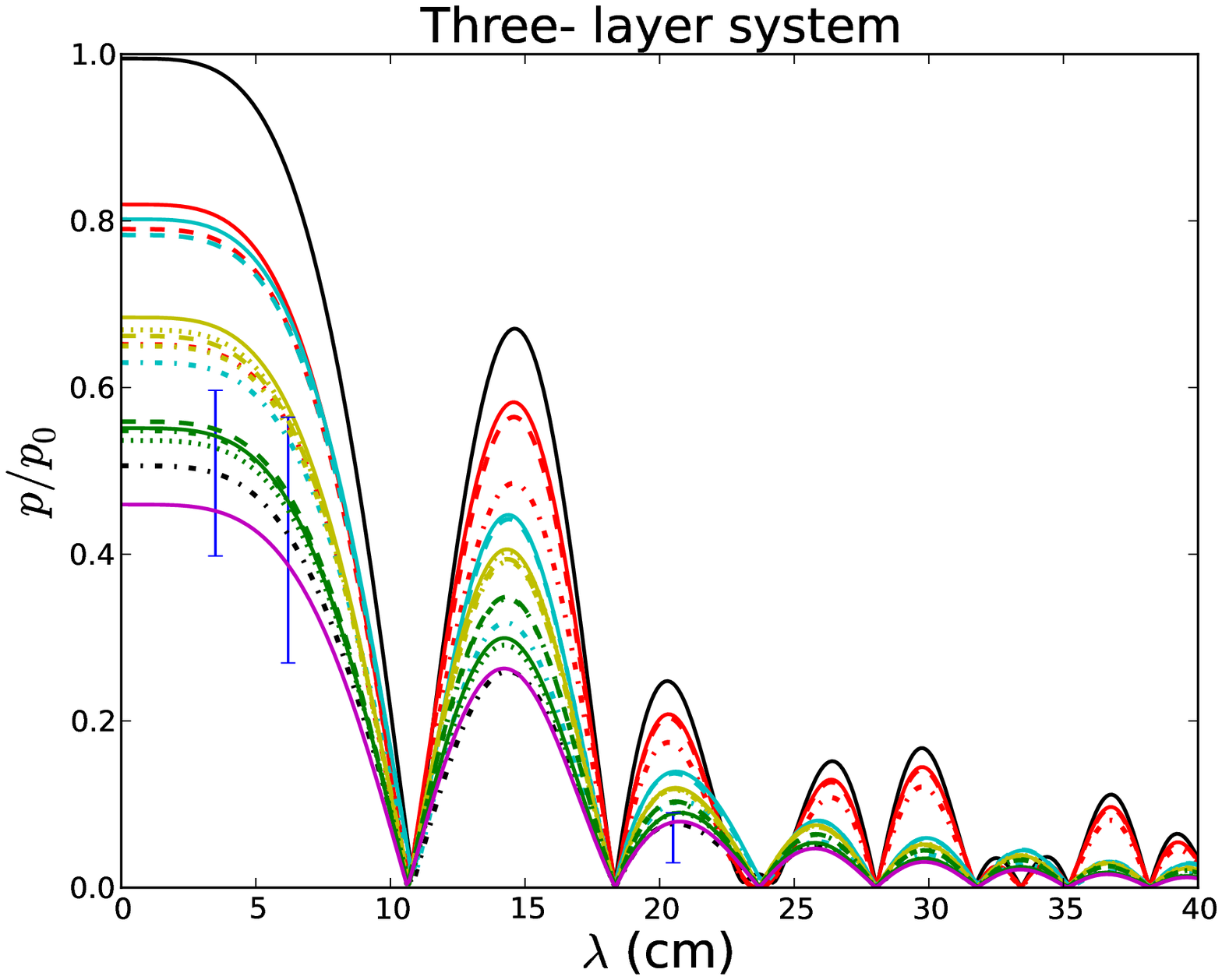}
\caption{Exactly the same model types and physical parameters as used in Fig.~\ref{Fig1_2layers} above but now for a three-layer system.}
\label{Fig2_3layers}
\end{figure}

\begin{table}
\caption{Model settings for Figs.~\ref{Fig1_2layers}~- \ref{Fig3_1layers} based on regular and turbulent magnetic field configurations in the disk and halo.}	
\label{table:1}
\begin{center}
\begin{tabular}{l c c c c c c }
\hline 
\hline 
\rule{0pt}{2.5ex} 
 & \multicolumn{3}{ c }{Disk} &  \multicolumn{3}{ c }{Halo} \\ 
\hline 
\rule{0pt}{2.5ex} 
 & Reg. & Iso. & Aniso. & Reg. & Iso. & Aniso. \\ 
\hline 
DH					&$\checkmark$&	&	&$\checkmark$&  	&	\\ 
DIH					&$\checkmark$&$\checkmark$&		&$\checkmark$&	&	\\
DAH					&$\checkmark$&	&$\checkmark$&$\checkmark$&		& 	\\	
DAIH					&$\checkmark$&$\checkmark$&$\checkmark$&$\checkmark$&	& 	\\
DHI					&$\checkmark$&	&	&$\checkmark$&$\checkmark$&		\\
DHA					&$\checkmark$&	&	&$\checkmark$&	&$\checkmark$	\\
DHAI					&$\checkmark$&	&	&$\checkmark$&$\checkmark$&$\checkmark$	\\
DIHI					&$\checkmark$&$\checkmark$&		&$\checkmark$&$\checkmark$&	\\
DIHI $\blacksquare$ 			&$\checkmark$&$\checkmark$& 	&$\checkmark$&$\checkmark$& 	\\
DIHA					&$\checkmark$&$\checkmark$& 	&$\checkmark$&	&$\checkmark$	\\
DAHI					&$\checkmark$&	&$\checkmark$&$\checkmark$&$\checkmark$&	\\
DAHA					&$\checkmark$&	&$\checkmark$&$\checkmark$&	&$\checkmark$	\\	
DIHAI					&$\checkmark$&$\checkmark$& 	&$\checkmark$&$\checkmark$&$\checkmark$	\\
DAHAI					&$\checkmark$&	&$\checkmark$&$\checkmark$&$\checkmark$&$\checkmark$	\\
DAIHI					&$\checkmark$&$\checkmark$&$\checkmark$&$\checkmark$&$\checkmark$&	\\
DAIHA					&$\checkmark$&$\checkmark$&$\checkmark$&$\checkmark$&	&$\checkmark$	\\
DAIHAI					&$\checkmark$&$\checkmark$&$\checkmark$&$\checkmark$&$\checkmark$&$\checkmark$	\\
\hline
D					&$\checkmark$&	&	&	&  	&	\\ 
DI 		  			&$\checkmark$&$\checkmark$&		&	&	& 	\\	
DI $\bigstar$  			&$\checkmark$&$\checkmark$&		&	&	& 	\\	
DI $\blacksquare$ $\bigstar$		&$\checkmark$&$\checkmark$&		&	&	&	\\ 
DA   					&$\checkmark$&	&$\checkmark$&	&	&	\\
DA $\bigstar$  			&$\checkmark$&	&$\checkmark$&	&	&	\\
DAI 					&$\checkmark$&$\checkmark$&$\checkmark$&	&	&	\\
DAI $\bigstar$ 			&$\checkmark$&$\checkmark$&$\checkmark$&	&	&	\\
DhI 					&$\checkmark$&		&	&		&$\checkmark$&	\\	
DIhI					&$\checkmark$&$\checkmark$&		&	&$\checkmark$&	\\
DIhI $\blacksquare$ 			&$\checkmark$&$\checkmark$& 		&	&$\checkmark$& 	\\
DIhI $\bigstar$ 			&$\checkmark$&$\checkmark$& 	&	&$\checkmark$& 	\\	
DIhI $\blacksquare$ $\bigstar$  	&$\checkmark$&$\checkmark$& 	&	&$\checkmark$& 	\\	
DAhI					&$\checkmark$&	&$\checkmark$&	&$\checkmark$&	\\
DAhI $\bigstar$			&$\checkmark$&	&$\checkmark$&	&$\checkmark$&	\\
DAIhI					&$\checkmark$&$\checkmark$&$\checkmark$&	&$\checkmark$&	\\
DAIhI $\bigstar$			&$\checkmark$&$\checkmark$&$\checkmark$&	&$\checkmark$&	\\
\hline
\hline	
\end{tabular} 
\tablefoot{ The three column headings below the principle headings of the `Disk' and `Halo' denote the regular, isotropic turbulent, and anisotropic turbulent magnetic fields. 
The rows contain a listing of all model types simulated with the following nomenclature: `D' denotes disk magnetic fields, `H' and `h' both denote halo magnetic fields, `I' and `A' are the isotropic and anisotropic turbulent magnetic fields, $\blacksquare$ represents the use of the $\lambda \, 3.5$ cm observations to gauge the wavelength-independent effects, and $\bigstar$ denotes the use of the generalized opaque-layer approximation to describe the contribution of internal Faraday dispersion (IFD) (in the disk) to depolarization, as detailed in Section~\ref{GaugeApproach}.  
Upper case letters `D' and `H' and the lower case `h' are used to distinguish between the presence or absence of a regular magnetic field in a given layer, respectively. 
The row ordering follows the model type order as in the legend of Figs.~\ref{Fig1_2layers} and~\ref{Fig2_3layers} for the top panel and that of Fig.~\ref{Fig3_1layers} for the bottom panel.   
}
\end{center}
\end{table}

Figures~\ref{Fig1_2layers} -~\ref{Fig2_3layers} constitute a snapshot, at a physically reasonable set of magnetic field values for the disk and halo, of all observationally motivated combinations that may be used to constrain field values for our example bin.
The particular magnetic fields underlying these figures involve a total regular disk and halo magnetic field strength of $5 \, \mu \mbox{G}$ each, an isotropic and anisotropic disk turbulent random field of $\sigma^2_{\text{I}} = \sigma^2_{\text{A}} = 10 \, \mu \mbox{G}$ for a total disk random field of about $14 \, \mu \mbox{G}$, and an isotropic and anisotropic halo turbulent random field of $\sigma^2_{\text{I}} = \sigma^2_{\text{A}} = 3 \, \mu \mbox{G}$ for a total halo random field of roughly $4 \, \mu \mbox{G}$. 
These total turbulent disk and halo magnetic field strengths are used to compute the disk and halo turbulent cell sizes of $40$ pc and $240$ pc, respectively.

\subsection{Generalized opaque-layer approximation} \label{GaugeApproach} 

We applied a generalized version of an approach, which was used by \citet{berk} to provide an approximate description to IFD, in order to predict depolarization values at the three observing wavelengths for M51 and test a method for parametrizing the depolarization, which is most significant at the $\lambda \, 20.5$ cm observing wavelength. The  opaque-layer approximation was defined by \citet{sokol}.
It assumes a thermal disk with uniform scale height $h_{th}$, a synchrotron disk with a wavelength-dependent, uniform scale height $h_{syn} $, and a thermal halo. Since $h_{syn} > h_{th}$, there is a narrow layer of synchrotron emission extending into the thermal halo. With the assumption that only the nearest part of the synchrotron emitting layer is visible due to depolarization, \citet{berk} estimate the contributions to the rotation measure from the disk and from the halo as $\mbox{RM} = \xi_d \, \mbox{RM}_d + \xi_h \, \mbox{RM}_h$, where ($\xi_d, \xi_h$) parametrize the disk and halo fractional RM contribution to the total observed RM. 
The $\xi$ parameters depend on the scale heights of the synchrotron disk and of both the thermal disk and halo and on the relative depolarization between the different observing wavelengths. There may be a variation with radius as well. In particular, the $\xi$ parameter values at $\lambda\lambda \, 3.5,6.2$ cm are close to unity, which implies that there is hardly any change to the actual Faraday depth at these two lower wavelengths.

\begin{figure}
\centering
\includegraphics[width=\hsize]{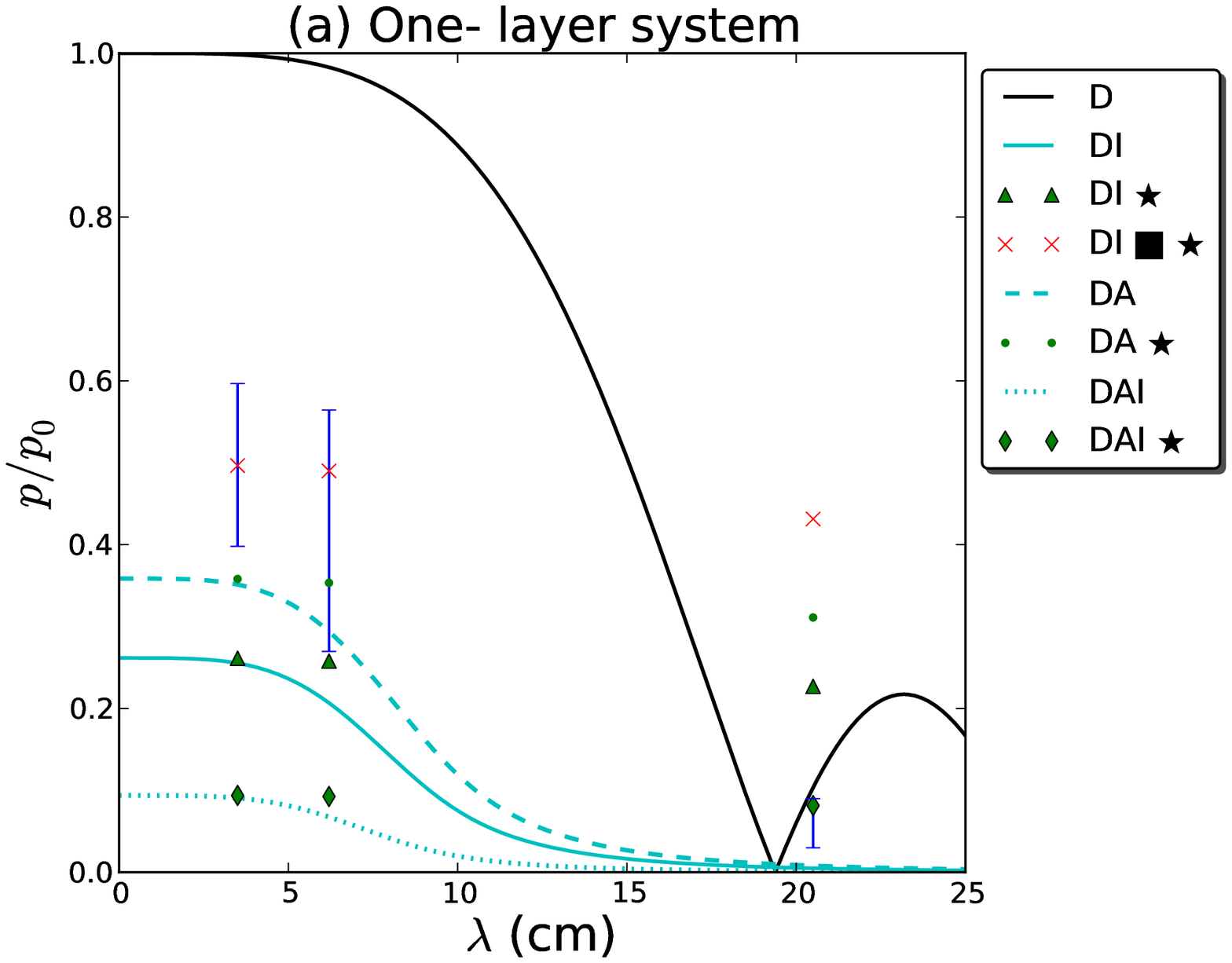} 
\includegraphics[width=\hsize]{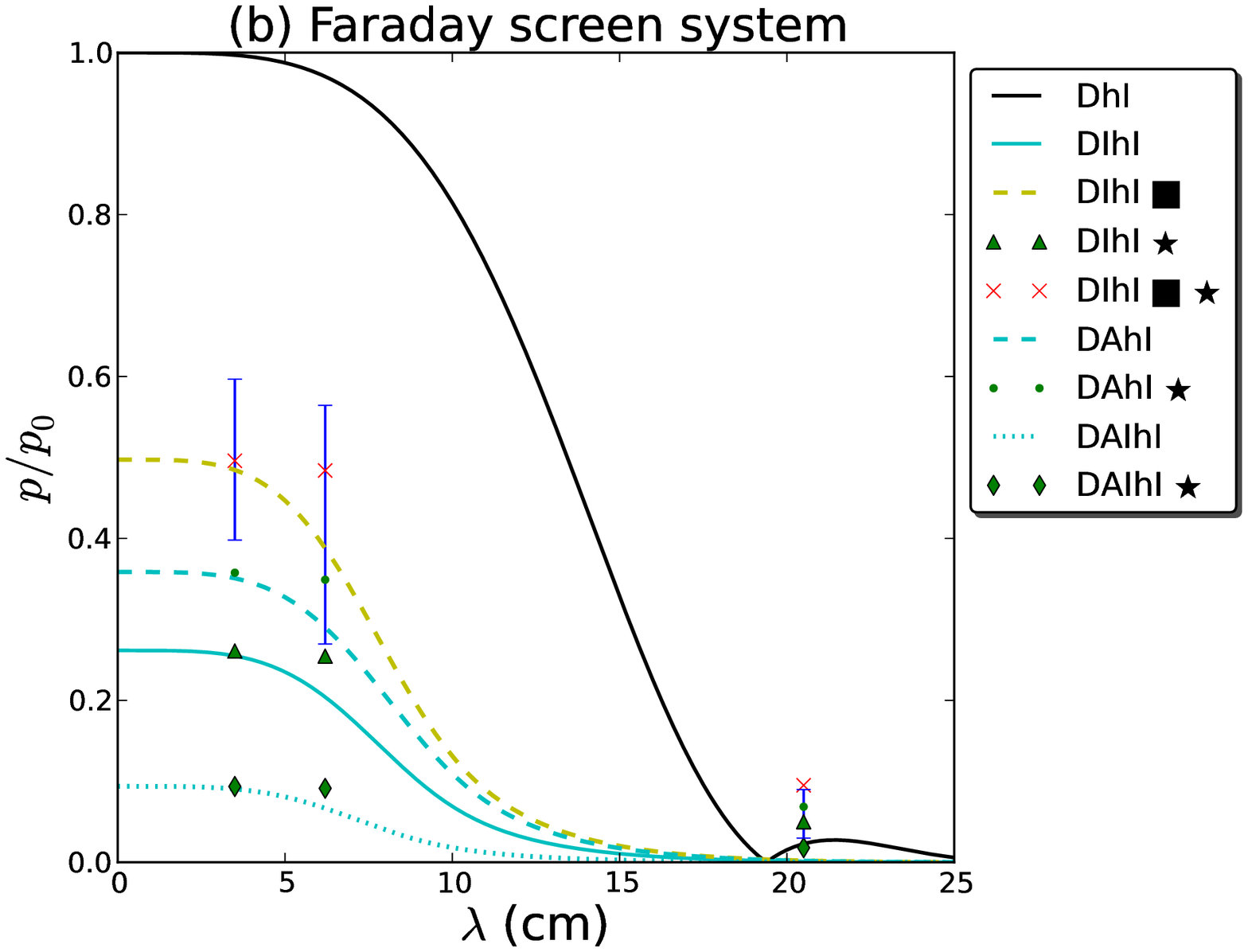} 
\caption{Normalized degree of polarization as a function of wavelength with the same physical parameters and nomenclature as in Figs.~\ref{Fig1_2layers} and~\ref{Fig2_3layers}. (a) One- layer system with a synchrotron emitting and Faraday rotating disk only. (b) The disk as in (a) but now with a halo that is only Faraday rotating.}
\label{Fig3_1layers}
\end{figure}

\citet{fletcher11} used the opaque-layer approximation to suppress Faraday rotation by the disk at the longest observing wavelength, while both the disk and halo Faraday rotate the emission at the shorter pair of observing wavelengths.
As we are dealing here with a Faraday screen system, we implement either of Eqs.~\eqref{Faradayscreen_sigma_halo_depol_terms_only} or~\eqref{Faradayscreen_depol_terms_only} and substitute the Faraday depth $R$ in Eq.~\eqref{Rsubi} by the RM values from \citet{berk}. To determine the depolarization as predicted by this approximation at the observing wavelengths, the scale heights of the synchrotron disk and of both the thermal disk and halo are used from \citet{berk}, but the relative depolarization are determined from the \citet{fletcher11} data. 
The generalized opaque-layer approximation may be combined with the assumption that all wavelength-independent depolarization effects are calibrated by observations of polarization at the lowest observing wavelength of $\lambda \, 3.5 \mbox{cm}$ \citep{berk}. 
Comparing Fig.~\ref{Fig3_1layers}(a) with Fig.~\ref{Fig3_1layers}(b) indicates that the presence of a turbulent magnetic field in the halo is required together with both the wavelength-independent gauge and opaque-layer approximation in order to have the best chance of fitting the data for the physically plausible regular and turbulent magnetic fields examined for the disk and halo.

\section{Discussion and conclusions} \label{Discussion}

The effectiveness of the method in generating a range of models for the diffuse ISM in M51, in terms of the number of magneto-ionic layers modeled and type of magnetic field species occurring in the disk and halo, is illustrated in Figs.~\ref{Fig1_2layers} - \ref{Fig3_1layers} for our example bin. With typical parameter values as in Table 1, one can immediately rule out models with regular fields only in the disk or in the disk and halo, in agreement with ubiquitous observations of turbulent magnetic fields in spiral galaxies. 

Even though the modeled magnetic field strengths can be varied for individual models in order to match the data values, the variation in the degree of polarization predicted by the range of models is much greater than the errors in the observed degree of polarization. This gives confidence that observations like these can indeed be used to rule out at least some of the
depolarization models. 

These models contain many potentially free parameters, which will mean the optimum solutions will be degenerate, however many of the parameters, specifically those in Table~\ref{table:2}, can be constrained using prior studies. The remaining free parameters are the regular field strengths and isotropic and anisotropic turbulent field strengths, both in the disk and halo. 

For these values to be well determined, a sufficient number of data points are needed. For the data from \citet{fletcher11}, containing only three wavelengths, data in one bin only (as shown in Figs.~\ref{Fig1_2layers} - \ref{Fig3_1layers}) cannot constrain the magnetic field strengths sufficiently. However, some additional assumptions about these field strengths can break the degeneracy. For example, we show in Paper II that the assumption of magnetic field strengths being independent of azimuth provides enough constraints to determine the regular and turbulent magnetic field strengths. With the broadband capabilities of most current radio interferometers, these depolarization curves can be sampled extremely well in wavelength space, with higher sensitivity, thus allowing actual tracing of these depolarization curves.

Throughout the paper, we have assumed a $p_0$ of $70 \%$ corresponding to the theoretical injection spectrum for electrons accelerated in supernova remnants ($\alpha_{syn} = -0.5$), as representative of the synchrotron spectral index $\alpha_{syn}$ in the spiral arms of M51 \citep{fletcher11}. 
For realistic, optically thin astrophysical plasmas, such as disks and halos of galaxies, $p_0$ ranges from $60\% to 80\%$ \citep[Section 3.3]{ginzburg_syrovatskii}.
\citet{fletcher11} estimated a constant $p_0$ of $76 \%$ across M51 ($\alpha_{syn} = -1.1$) but observed variations in this value.
This would imply that our current reported values of $p/p_0$ at the three observing wavelengths are, on the whole, $8 \%$ higher than the expected polarization value. 
However, this overestimate is small compared to the 
$20 \% to 50 \%$ margin of error in the observations at each of the three observing wavelengths.
With better data having errors of only a few percent, the spectral index maps of \citet[Fig.7]{fletcher11} would have to be binned in the same way as the polarization maps, and the resulting $p_0$ value per bin would have to be used.

In general, an anisotropic field has a higher degree of polarization than an isotropic field when comparing fields of equal total strength. The greater the anisotropic $\alpha$ term, the higher the polarization. 
The anisotropic and isotropic turbulent components are presently modeled as yielding two independent depolarization contributions in separate parts of the medium with the strength of IFD determined by the total turbulent field. 
The next step in modeling would be to include an anisotropic random component in the complete medium and to modify $\sigma_{RM}$ to reflect an angular dependence in the presence of the anisotropic field.
Moreover, if a non-constant spectral index were to be considered, then the effect of (spatial) spectral variation on polarization would have to be accounted for \citep{bJburn,Gardner66}.
The purpose of this work is not to arrive at exact equations for depolarization that are able to incorporate the effects of a greater number of depolarization mechanisms but rather to offer a useful approach to modeling and deducing certain physical parameters of the magneto-ionic medium being analyzed from its polarized emission. 

 We have shown that various models of depolarization in the disk and halo give widely differing predictions for depolarization at various wavelengths, making them a useful tool for estimating the disk's and halo's regular and turbulent magnetic fields.
 Our method incorporates depolarizing effects in the disk and halo directly and allows for simultaneous depolarization contributions from DFR and IFD. 
 We also treated depolarization due to anisotropic turbulent fields, albeit with simplifying assumptions described earlier.
 Modeling the disk and halo as both a two- and three-layer synchrotron emitting and Faraday rotating system allows for the depolarization contribution of the far side of the halo to be examined.  
 A model of the galaxy's regular field is required as an input. 
 The multilayer modeling approach with the inclusion of anisotropic turbulent magnetic fields is found to be a more suitable prescription for the data.
 For the two-layer system where the halo functions as a Faraday screen, the opaque-layer approximation may work under certain circumstances, but not always. 
This may be due to oversimplification of the model and/or a lack of a synchrotron halo in the model.

 Our method is more robust than the opaque-layer approximation because it is based on more fundamental physical parameters of the galaxy rather than on a wavelength-dependent synchrotron scale height parametrization. We modeled the effects of wavelength-independent and wavelength-dependent depolarization directly, which allowed for a statistical comparison with the polarization maps at the observing wavelengths.
The different models provide different enough results that existing multiwavelength observations of nearly face-on galaxies can distinguish between them.

\begin{acknowledgements}
CS would like to express his gratitude to Huub R\"{o}ttgering for his generous time and most supportive supervision. 
CS and MH acknowledge the support of research program 639.042.915, which is partly financed by the Netherlands Organization for Scientific Research (NWO).
AF and AS thank the Leverhulme trust for financial support under grant RPG-097.
We are grateful to the anonymous referee for the prompt and most helpful comments and suggestions for strengthening the paper.  
\end{acknowledgements}

\begin{appendix} \label{AppendixA} 

\section{Derivation of wavelength-independent depolarization equations for standard and equipartition scalings of emissivity} \label{derivofpre_27_Sokol} 
 
 We derive the results of \citet{sokol} for wavelength-independent depolarization to explicitly show how the corresponding equations arise for two different scalings of emissivity along with the independence of the intrinsic polarization angle from these scalings. We also correct two slight errors in the formula for emissivity given in \citet{sokol} for the case of energy equipartition.

 For a total magnetic field that is purely a regular (mean) field, $\boldsymbol{B} = \overline{\boldsymbol{B}}$, the complex \emph{intrinsic} (hence wavelength-independent) polarization $\mathcal{P}_{0i}$ per layer $i$ is given by 
 \begin{equation} \label{intrinsic_pol_reg}
 \mathcal{P}_{0i} = p_0 \, \exp\left({2 \imath \psi_{0i}}\right),
 \end{equation}
 where $p_0$ is the intrinsic degree of polarization, and $\psi_{0i}$ is the initial polarization angle per layer $i$.

 In the presence of a turbulent magnetic field $\boldsymbol{b}$, the total field becomes $\boldsymbol{B} = \overline{\boldsymbol{B}} \, + \, \boldsymbol{b}$ and, together with a sufficiently large number of correlation cells encompassed by the telescope beam cylinder, the volume average in the synchrotron emitting source becomes equal to the ensemble average via the ergodic hypothesis, and $\mathcal{P}_{0i}$ is modified from the above Eq.~\eqref{intrinsic_pol_reg} to what is given by Eq.~\eqref{Pnaught_new}
 \begin{equation} \label{Pnaught_new_again}
 \left\langle\mathcal{P}_{0i}\right\rangle = p_0 \frac{\left \langle \varepsilon_i \, \exp\left({2 \imath \, \psi_{0i}}\right) \right \rangle}{\left \langle \varepsilon_i \right \rangle}, 
 \end{equation}
 where $\varepsilon_i$ is the synchrotron emissivity and $\langle \ldots \rangle$ denotes ensemble averaging. This expectation value entails computing various moments of the total magnetic field components.
 
 To determine how the intrinsic polarization value $p_0$ has been modified, in effect, by the presence of turbulence to a layer dependent value $p_{0i}$ ($p_0$ itself remains constant and equal to $0.7$), the quantity $\left| \left\langle\mathcal{P}_{0i} \right\rangle\right| /p_{0}$ has to be evaluated. 

 Assuming that the total magnetic field is a random Gaussian variable, a Taylor expansion of the moment-generating function $M$ for a normal or Gaussian distributed random variable $X$ defined as
 \begin{equation} \label{moment_gen_fcn}
 M_X(s) = \exp \, \left(s \, \mu \, + \, \tfrac{1}{2} \, \sigma^2 \, s^2 \right) 
 \end{equation}
is performed about $s=0$ to yield equations for $m_n$, the $n^{th}$ moment of $M_X$, at each $n^{th}$ derivative of the function. Therefore, $m_n$ is to be identified with $\left\langle X^n \right\rangle$.
 
 The explicit computation of moments of $M_X$ in Eq.~\eqref{moment_gen_fcn} may be explained as follows. For a given layer $i$, whether disk or halo, substitute $X$ by the successive components of the total field $\boldsymbol{B}$, which is a random variable because it is the sum of a regular and random variable, and replace $s$ with appropriate instances of the three spatial directions in Cartesian coordinates $x,y,z$. Then identify $\mu$ as an instance of the mean $\overline B_{x,y,z}$ and $\sigma^2$ as an instance of the variance\footnote{The variance of a complex random variable $X$ is given by $\sigma^2_X = \left \langle \, \left( X \, - \, \left \langle \, X \, \right \rangle \right) \, \left(X^{*} \, - \, \left \langle \, X^{*} \, \right\rangle \right) \, \right \rangle = \left \langle X \, X^{*} \right \rangle \, - \, \left \langle X \right \rangle \, \left \langle X^{*} \right \rangle$, where the asterisk denotes the complex conjugate.} $\sigma^2_{x,y,z}$ of the corresponding components of $\boldsymbol{b}$. 

 For completeness, the first through fourth moments are 
 \begin{align*} 
 m_1 &= \mu, \nonumber \\
 m_2 & = \mu^2 \, + \sigma^2, \nonumber \\
 m_3 & = \mu^3 \, + \, 3 \, \mu \, \sigma^2, \nonumber \\
 m_4 & = \mu^4 \, + 3 \, \sigma^4 \, + \, 6 \, \mu^2 \, \sigma^2.  \nonumber
 \end{align*}
For the case of a purely random field, $\mu = 0$ leaving only the even (central) moments $m_2$ and $m_4$. For the case of a purely regular field, $\sigma = 0$ and the four moments simply reduce to the first through fourth powers of the mean field. 

 Assuming that the emissivity per layer $i$ scales as
 \begin{equation} \label{reg_scaling}
 \varepsilon_i = c \, B^2_{\perp i}, 
 \end{equation} 
 the complex emissivity is, therefore, given by
 \begin{equation} \label{reg_complex_angle}
 \varepsilon_i \, \exp  \, (2 \imath \, \psi_{0i}) \, = \, c \, (B^2_{xi} - B^2_{yi} + 2 \imath \, B_{xi} \, B_{yi}), 
 \end{equation}
 where $B_{\perp i} = B_{xi} + \imath B_{yi}$, $B^2_{\perp i} = \left|B_{\perp i}\right|^2 = B^2_{xi} + B^2_{yi}$, and $c$ is a constant depending on the number density of relativistic cosmic ray electrons $n_{\text{cr}}$. Taking the square of each of the two equivalent representations of a complex number $z$ as given by $R\exp\left(\imath \theta\right) = z = x \,+\, \imath y$, with $R = \left|x + \imath y \right|$ and $\tan\theta = \text{Im}\left(z\right)/\,\text{Re}\left(z\right) = y/x$ and identifying the coordinates $x,y$ with $B_{xi},B_{yi}$ may serve as an aid in arriving at Eq.~\eqref{reg_complex_angle}. 

The absolute value of Eq.~\eqref{Pnaught_new_again} with the emissivity scaling of Eq.~\eqref{reg_scaling} therefore yields the following equation for the wavelength-independent depolarization as in Eq.~\eqref{gen_form} and as in Eq. (19) of \citet{sokol}. 
\begin{equation*} \label{sokol_19} 
 \frac{\left| \left\langle \mathcal{P}_{0i} \right\rangle \right|}{p_0} = \left\{ \frac{\left \lbrack \left( \overline B^2_x - \overline B^2_y + \sigma^2_x - \sigma^2_y \right)^{2} + 4 \overline B^2_x  \overline B^2_y \right \rbrack^{1/2}}{\overline{B^2_\perp}} \right\}_i, 
 \end{equation*}
 where $\overline{B}^2_\perp = \overline B^2_x \,+ \, \overline B^2_y$, \, $\overline{B^2_\perp} = \overline B^2_\perp \, + \, \sigma^2_x \, + \, \sigma^2_y $.

The intrinsic polarization angle is also modified and obtained from the ratio of imaginary to real parts of the expectation value of the complex emissivity via $\tan \left(2 \left\langle \psi_{0i} \right\rangle \right) = \text{Im}\left(\left\langle\text{Eq.~\eqref{reg_complex_angle}}\right\rangle\right)\, /  \,\text{Re}\left(\left\langle\text{Eq.~\eqref{reg_complex_angle}}\right\rangle\right)$ and is therefore given by
 \begin{equation} \label{psi_gen_again}
 \left \langle \psi_{0i} \right \rangle = \tfrac{1}{2}\pi \, + \, \tfrac{1}{2} \arctan \left( \frac{2 \overline B_x \overline B_y}{\overline B^2_x - \overline B^2_y + \sigma^2_x - \sigma^2_y} \right)_i 
 \end{equation}
as in Eq.~\eqref{psi_gen} without the sky-plane coordinate transformation term and as in Eq. (20) of \citet{sokol}.

With the energy equipartition and pressure equilibrium assumptions the cosmic ray number density scales as $n_{cr} \propto B^2$ if the energy densities of magnetic fields and cosmic rays are completely correlated, and the scaling of synchrotron emissivity with magnetic field becomes 
 \begin{equation} \label{equip_scaling}
 \varepsilon_i = C \, B^2_i \, B^2_{\perp i} 
 \end{equation}
with a certain constant $C$, therefore
 \begin{equation} \label{equip_complex_angle}
 \varepsilon_i \, \exp  \, (2 \imath \, \psi_{0i}) \, = \, C \, B^2_i \, (B^2_{xi} - B^2_{yi} + 2 \imath \, B_{xi} \, B_{yi}),
 \end{equation}
where $B^2_i = B^2_{xi} + B^2_{yi} + B^2_{zi}$. 
The intrinsic polarization angles are \emph{unaffected} by the rescaling of emissivity since the constant term $C B^2_i$ cancels out, exactly like the $c$ term, in arriving at Eq.~\eqref{psi_gen_again}. In addition to the first two moments, the third and fourth moments of the fields $B_k$ with $k = \{x,y,z\}$ in~\ref{equip_scaling},~\ref{equip_complex_angle} must be computed.

 Consequently, the absolute value of Eq.~\eqref{Pnaught_new_again} transforms to 
\begin{align} \label{pre_27_Sokol}
      \frac{\left| \left\langle \mathcal{P}_{0i} \right\rangle \right|}{p_0} = & \left[\overline{B^2} \, \overline{B_\perp^2} \, + \, 2 \left( \sigma^4_x + \sigma^4_y \right) \, + \, 4 \left( \overline B^2_x \, \sigma^2_x \, + \, \overline B^2_y \, \sigma^2_y \right) \right]^{-1} \nonumber \\
	& \times  \bigg \lbrace \Bigl[ \overline B^4_x - \overline B^4_y + 3 \left( \sigma^4_x - \sigma^4_y \right) + 6 \left( \overline B^2_x \, \sigma^2_x - \overline B^2_y \, \sigma^2_y \right)  \nonumber \\
	& + \overline{B^2_{||}} \left( \overline{B^2_x} - \overline{B^2_y} \right) \Bigr]^{2} + 4\overline B^2_x \overline B^2_y \left \lbrack \overline{B^2} + 3 \left( \sigma^2_x + \sigma^2_y\right) \right \rbrack^{2} \bigg \rbrace^{1/2},
 \end{align}
where the righthand side of the above equation is to be taken per individual layer $i$, disk or halo, $\overline{B^2_{||}} = \overline B^2_{||} + \sigma^2_{||} $ and $\overline{B^2} = \overline{B^2_\perp} + \overline{B^2_{||}} $. 
Isotropy is now given by $\sigma_x = \sigma_y = \sigma_{||} = \sigma $. 
The form of Eq.~\eqref{pre_27_Sokol} would then imply the corresponding modification in Eqs.~\eqref{gen_form} - \eqref{iso_and_aniso}. The simple multiplicative relationship between the wavelength-dependent and wavelength-independent terms as represented in Eq.~\eqref{complex_Pol_turb_new_final} would continue to hold only if no dependence on the line-of-sight coordinate arose.

\section{Symmetries and equation properties} \label{eqndiscussion}

Symmetry considerations are appropriate for discussion in the context of depolarization. 
Layer \emph{ordering} and line-of-sight magnetic field \emph{reversal} are two distinct symmetries that arise in our modeling.
Layer ordering involves a reflection of the physical system or the placement of the observer at the opposite end of the originally oriented system. For a two-layer medium this simply involves an exchange of the index $i$ that also causes $\Delta \psi_{dh} \rightarrow -\Delta \psi_{dh}$.
For a three-layer system, with identical far and near sides of the halo, reflection is automatically satisfied. 
For magnetic field reversal along the line of sight, only the direction of the line-of-sight \emph{regular} field has to be reversed $\overline{B}_z \rightarrow -\overline{B}_z$, in all layers at once, since a change of sign for turbulent fields has no affect on polarization. 

With only DFR present, the equation for depolarization in a two-layer system, given by Eq.~\eqref{2layer_dfr}, indicates that the presence of the $\Delta\psi$ term breaks each of the ordering and reversal symmetries but that symmetry is preserved only if both layer ordering and field reversal are applied \emph{simultaneously}. A three-layer system remains invariant under field reversal as apparent from Eq.~\eqref{3layer_dfr}. 

IFD occurring with DFR changes the previously encountered symmetry properties for DFR alone in terms of layer ordering and field reversal for a two- and three-layer system. 
In particular, it is always the cross terms (which mix the layers) that determine these symmetries. A two-layer system given by Eq.~\eqref{directsum_2layer} remains invariant under the line-of-sight regular magnetic field sign inversion only when the disk and halo intrinsic polarization angles are equal ($\Delta \psi_{dh} = 0$) just as for the two-layer system with DFR alone. However, the IFD `carrier' $\sigma_{RM}$ terms break the previously achieved layer ordering symmetry so that the two-layer system becomes sensitive to whether the far or near side of the halo is switched on alongside the disk. 
For a three-layer system given by Eq.~\eqref{directsum_3layer}, the presence of IFD now imposes the extra condition that the disk and halo intrinsic polarization angles must be equal in order to have the field reversal symmetry as for the two-layer system. 
For a Faraday screen system, Eq.~\eqref{Faradayscreen_sigma_halo_depol_terms_only} remains \emph{symmetric} under the reversal of the total magnetic field direction along the line of sight $B_z \rightarrow -B_z$. When the symmetries are broken, the amplitude and period are only slightly affected for our example bin.
Both of the three-layer Eqs.~\eqref{directsum_3layer} and~\eqref{3layer_dfr} contain a non-trivial $\left(1+ \cos \left(2 \left(R_d + R_h \right) \, \lambda^2 \right) \right)$ term that contains the combined actions of the disk and near halo regular fields and arises from the near and far sides of the halo being set identically equal.

\end{appendix}

\bibliographystyle{aa} 

\end{document}